\documentclass[fleqn,usenatbib]{mnras}

\usepackage[T1]{fontenc}

\DeclareRobustCommand{\VAN}[3]{#2}
\let\VANthebibliography\thebibliography
\def\thebibliography{\DeclareRobustCommand{\VAN}[3]{##3}\VANthebibliography}

\usepackage{graphicx}	
\usepackage{amsmath}	
\usepackage{amssymb}	

\usepackage{color}
\definecolor{darkgreen}{RGB}{0,142,128}

\newcommand{\moddAS}[1]{{#1}}
\newcommand{\modAS}[1]{{#1}}
\newcommand{\refAS}[1]{{#1}}

\usepackage{multirow}

\newcommand{\dr}{\partial_r}

\newcommand{\dth}{\partial_\theta}
\newcommand{\dphi}{\partial_\varphi}

\newcommand{\bnab}{\boldsymbol{\nabla}}
\newcommand{\rot}{\bnab\times}
\newcommand{\Div}{\bnab\cdot}

\newcommand{\er}{\mathbf{e}_r}
\newcommand{\ethe}{\mathbf{e}_\theta}
\newcommand{\ephi}{\mathbf{e}_\varphi}

\newcommand{\gradperp}{\bnab_\perp}

\newcommand{\Rlm}[1]{\mathbf{R}^{m_{#1}}_{l_{#1}}}
\newcommand{\Slm}[1]{\mathbf{S}^{m_{#1}}_{l_{#1}}}
\newcommand{\Tlm}[1]{\mathbf{T}^{m_{#1}}_{l_{#1}}}
\newcommand{\Ylm}[1]{Y_{l_{#1}}^{m_{#1}}}

\title[MOVES V. Modelling star-planet magnetic interactions of HD 189733 in August 2013]{MOVES V. Modelling star-planet magnetic interactions of HD 189733}

\author[A. Strugarek \textit{et al.}]{
A. Strugarek,$^{1}$\thanks{E-mail: antoine.strugarek@cea.fr}
R. Fares,$^{2}$
V. Bourrier,$^{3}$
A. S.  Brun,$^{1}$
V. R\'eville,$^{4}$
\newauthor
T. Amari,$^{5}$
Ch. Helling,$^{6,7}$ 
M. Jardine,$^{8}$
J. Llama,$^{9}$
\newauthor
C. Moutou,$^{4}$
A. A. Vidotto,$^{10}$
P. J. Wheatley,$^{11,12}$
P. Zarka$^{13}$
\\
$^{1}$D\'epartement d'Astrophysique/AIM, CEA/IRFU, CNRS/INSU, Univ. Paris-Saclay, Univ. de Paris, 91191 Gif-sur-Yvette, France\\
$^{2}$Physics Department, United Arab Emirates University, P.O. Box 15551, Al-Ain, United Arab Emirates\\
$^{3}$Observatoire Astronomique de l'Universit\'e de Gen\`eve, Chemin Pegasi 51b, 1290, Versoix, Switzerland\\
$^{4}$IRAP, Universit\'e Toulouse III - Paul Sabatier, CNRS, CNES, 14 avenue E. Belin, 31400 Toulouse, France\\
$^{5}$Centre de Physique Th\'eorique, CNRS, Ecole Polytechnique, IP Paris, F-91128 Palaiseau, France\\
$^{6}$ Space Research Institute, Austrian Academy of Sciences, Schmiedlstrasse 6, A-8042 Graz, Austria\\
$^{7}$ TU Graz, Fakult\"at f\"ur Mathematik, Physik und Geod\"asie, Petersgasse 16, 8010 Graz, Austria\\
$^{8}$SUPA, School of Physics and Astronomy, North Haugh, St Andrews, Fife KY16 9SS, UK\\
$^{9}$Lowell Observatory, 1400 W. Mars Hill Rd. Flagstaff. AZ. 86001. USA\\
$^{10}$Leiden Observatory, Leiden University, PO Box 9513, 2300 RA Leiden, The Netherlands\\
$^{11}$ Department of Physics, University of Warwick, Gibbet Hill Road, Coventry CV4 7AL, UK \\
$^{12}$ Centre for Exoplanets and Habitability, University of Warwick, Gibbet Hill Road, Coventry CV4 7AL, UK \\ 
$^{13}$LESIA, UMR CNRS 8109, Observatoire de Paris, 92195 MEUDON, France\\
}

\date{Accepted XXX. Received YYY; in original form ZZZ}

\pubyear{2022}

\usepackage{newtxtext,newtxmath}

\begin{document}
\label{firstpage}
\pagerange{\pageref{firstpage}--\pageref{lastpage}}
\maketitle

\begin{abstract}
Magnetic interactions between stars and close-in planets may lead to a detectable signal on the stellar disk. HD 189733 is one of the key exosystems thought to harbor magnetic interactions, which may have been detected in August 2013. We present a set of twelve wind models at that period, covering the possible coronal states and coronal topologies of HD 189733 at that time. We assess the power available for the magnetic interaction and predict its temporal modulation. By comparing the predicted signal with the observed signal, we find that some models could be compatible with an interpretation based on star-planet magnetic interactions. \moddAS{We also find that the observed signal can be explained only with a stretch-and-break interaction mechanism, while that the Alfv\'en wings scenario cannot deliver enough power.} We finally demonstrate \refAS{that the past observational cadence of HD 189733 leads to a detection rate of only between 12 to 23\%, which could explain why star-planet interactions have been hard to detect in past campagins. We conclude} that the firm confirmation of their detection will require dedicated spectroscopic observations covering densely the orbital and rotation period, combined with scarcer spectropolarimetric observations to assess the concomitant large-scale magnetic topology of the star.    
\end{abstract}




\section{Introduction}
\label{sec:intro}

Planets on close-in orbit around their host star are subject to complex interactions \citep{Lanza2018b}: tidal interactions \citep{Mathis2018}, ionization and atmospheric escape \citep[\textit{e.g.}][]{Owen2019,Gronoff2020}, interactions with transients stellar events \citep{Alvarado-Gomez2020,Varela2021}, and direct magnetic interactions \citep{Saur2017,Strugarek2018a}. The detection of such interactions \citep{Shkolnik2018} can provide unique constraints to characterize the secular evolution of star-planet systems \citep[\textit{e.g.}][]{Ahuir2021a,Lazovik2021}, planetary magnetic field \citep[\textit{e.g.}][]{Cauley2018,Turner2021,Vedantham2020}, the atmospheric state of irradiated planets \citep[\textit{e.g.}][]{Bourrier2018,Bourrier2018b}, and even stellar wind properties \citep[\textit{e.g.}][]{Vidotto2017,Carolan2021}. 

Here, we will focus on the case of magnetic interactions. Two main detection techniques have been proposed so far to detect star-planet magnetic interactions (SPMIs). 

First, \citet{Cuntz2000} argued that the presence of a close-in planet could increase and modulate the activity of the host star due to direct magnetic connection. \citet{Shkolnik2003} first reported hints of this interaction in the variability of the Ca II H\& K bands of HD 179949. In this case, the modulation was observed with the same period as the orbital period of the hot Jupiter HD 179949b. Such detections have then been reported on other stars hosting hot Jupiters (see review of \citealt{Shkolnik2018}). Such signals are observed for some specific epochs, and was not found at others \citep[\textit{e.g.}][]{Shkolnik2008,Cauley2018}. If this signal originated from a magnetic interaction, such an on/off nature is expected due to the variability of the stellar magnetic field \citep{Moutou2007,Fares2017a}
mediating the magnetic interaction \citep{Cranmer2007,Strugarek2018a}. \moddAS{In addition, other tracers of magnetic activity can also be used to corroborate this intepretation. For instance, \citet{Gao2021} unveiled that specific longitudes of HD 189733 were particularly active in X-rays, which could be related to a star-planet magnetic connection operating in the system.} 

Second, SPMIs can also in principle be detected from radio emissions \citep{Zarka2007}. In that case, the emission can originate from two sources. On one side, the motion of the planet within the magnetosphere of the star is analogous to the motion of the satellites of Jupiter within its magnetosphere. The associated radio emissions would then originate from a location in the lower stellar atmosphere where accelerated electron populations would be unstable due to a cyclotron-Maser instability \citep[\textit{e.g.}][]{Zarka2004a}. On the other side, the radio emission could come from the planetary magnetosphere itself. In that case, the emission is directly correlated with the strength of the planetary magnetic field. Coherent radio emission that could be associated with SPMI have been recently reported by \citet{Vedantham2020,Turner2021,Callingham2021}. \moddAS{If the radio signal shows a modulation with the orbital period of a known planet, or with the rotation period of the star, it could be furthermore possible to disambiguate the source of the detected radio emission (stellar or planetary magnetosphere, see \textit{e.g.} \citealt{Fares2010a,Hess2011,Kavanagh2021})}.

In all cases, the stellar magnetic field plays a dominant role in shaping the SPMI signal. Indeed, the stellar wind close to the star is structured by the topology of the stellar magnetic field, and this topology acts as a guide for the SPMI signal from the planet to the star. Spectropolarimetric observations of stars allow the characterization of their large scale magnetic field thanks to the Zeeman-Doppler Imaging (ZDI) technique \citep{Donati2009}. Based on such magnetic maps, models of stellar coronae can predict the large-scale architecture of the stellar atmosphere up to the orbit of planets in close-in orbits (see \textit{e.g.} \citealt{Vidotto2018} and references therein). For instance, \citet{Strugarek2019} studied SPMIs in Kepler-78 and were able to show that they were likely not detectable in this system. The goal of this work is to assess the likelihood that SPMIs are at the origin of the signal detected in HD 189733 by \citet{Cauley2018,Cauley2019}. 

\citet{Cauley2018} carried a large study of HD 189733 over 6 different observational windows from June 2006 to July 2015. They analyzed the signal in the Ca II K band and removed a rotational modulation from it. In one epoch out of six (August 2013) they found a signal presenting a modulation close to the orbital period of the hot Jupiter HD 189733b. Such signal was interpreted as a SPMI signal in \citet{Cauley2019} to provide an estimate of the hypothetical magnetic field of HD 189733b. 

To assess the robustness of this interpretation, we present in this fifth paper of the MOVES collaboration a modelling effort of the wind of HD 189733 for August 2013. In the MOVES series of papers, the first paper by \citet{Fares2017a} was dedicated to five epochs of observations including August 2013. A magnetic map obtained with ZDI was derived and analyzed for each epoch. In the second paper of the series, \citet{Kavanagh2019} modelled the wind of HD 189733 for three epochs (June/July 2013, September 2013, and July 2015), focusing on the possibility to detect radio emissions originating from this exosystem. Paper III focused on the variable \refAS{X-ray} and UV environment of HD 189733 \citep{Bourrier2020}, and paper IV on the atmospheric composition of HD 189733b \citep{Barth2021}. Here we complement this series by using the ZDI map of HD 189733 in August 2013 to model its wind and by focusing on the manifestations of star-planet magnetic interactions.

We present in section \S\ref{sec:windmodelling} the modelling choices considered in this work. For August 2013, we have carried out 12 different polytropic wind models varying the coronal temperature and density assumed for HD 189733, as well as the magnetic field component used to deduce the 3D corona and wind. We present the resulting coronal 3D structure as well as the wind properties at the planetary orbit in \S\ref{sec:multiEpoch}. We then estimate the amplitude and temporal variability of SPMIs in \S\ref{sec:ExpSPMI}, and compare them to the signal detected by \citet{Cauley2019}. We find that only a few models can accommodate the SPMI interpretation of the observed CaII signal. We conclude that any further SPMI detection could be confirmed only with much more densely sampled observational campaigns better covering both the rotation period of HD 189733 and the orbital period of HD 189733b (see \S\ref{sec:conclusions}).

\section{Modelling the wind of HD 189733}
\label{sec:windmodelling}

Modelling the wind of HD 189733 based on observed Zeeman-Doppler Imaging maps requires several assumptions, which we detail in this section. The wind model used here is based on the Wind-Predict framework \citep{Reville2016a} leveraging the PLUTO code \citep{Mignone2007a}, which has been used for instance to model the corona and wind of Kepler-78 \citep{Strugarek2019}. The equations solved, and boundary conditions, can be found in details in \citet{Reville2015,Reville2016a}. \refAS{The numerical grid is the same as in \citet{Strugarek2019}, it is dense at the bottom of the domain and stretches away from the central star.} In this section, we discuss two specific aspects of this modelling. We first discuss the coronal parameters choices in \S\ref{sec:coroprop}, and then turn to the magnetic field extrapolation choices in \S\ref{sec:magExtrapol}. In addition, we recall the fundamental parameters of HD 189733 and its close-in planet HD 189733b in Table \ref{ta:HD189733prop}.

\subsection{Coronal properties and coronal modelling}
\label{sec:coroprop}

The modelling of stellar coronae and wind relies on several assumptions for the properties of the lower atmosphere of cool stars. In a previous work in this series, \citet{Kavanagh2019} modelled the corona of HD 189733 based on a polytropic wind model with the BATS-R-US code. A similar polytropic approach was also used to model the corona of Kepler-78 based on the PLUTO code \citep{Strugarek2019}. Here, we will make use of the latter model, and explore the sensitivity of the predicted star-planet magnetic interaction with respect to the wind model assumptions. 

Polytropic wind models rely on two main thermodynamic parameters at the base of the modelled domain: the coronal density $n_c$ and the coronal temperature $T_c$. For a given magnetic topology, and a given polytropic index $\gamma$, these two parameters control the mass loss rate of the stellar wind and the extent of the Alfv\'en surface. The latter correspond to the characteristic surface where the accelerating wind overcomes the local Alfv\'en speed \citep[for a review see][]{Strugarek2018a}. This surface plays a major role in star-planet magnetic interactions: planets orbiting outside the Alfv\'en surface cannot induce any magnetic interaction tracer on the stellar disk, while planets orbiting inside may do so. In \citet{Kavanagh2019}, the coronal density and temperature were chosen following the scaling laws of \citet{Johnstone2015a}. This leads to a high coronal temperature $T_c=2$ MK, and a relatively dense corona with $n_c=10^{10}$ cm$^{-3}$. These parameters give a mass loss rate of $\dot{M} \sim 3 \times 10^{-12} \, M_\odot$ yr$^{-1}$. Other choices can be made regarding these parameters. \citet{Ahuir2020} developed a methodology to estimate $n_c$ and $T_c$ from stellar parameters. In their approach, they used all observational data available to date (namely magnetic field detections with ZDI and Zeeman broadening, detected astrospheres and their associated mass loss, stellar X-ray flux, and the rotational distribution of stars in open clusters) to estimate realistic ranges of value for both parameters. Several scenarios are considered, and min/max scaling laws are provided for $T_c$ and $n_c$. Based on their approach and on the detected rotation period $P_{\rm rot} \simeq 12$ days \citep{Fares2010a}, we find that HD 189733 is an intermediate rotator with a Rossby number of $Ro \sim 0.3 Ro_\odot$ ($Ro_\odot$ is the solar Rossby number), and it is therefore rotating too slowly to be in the saturated regime. The minimum and maximum values allowed in their modelling for $n_c$ and $T_c$ are then given by the following scaling laws
\begin{align}
&\left\{
\begin{array}{ccc}
    n_c^{\rm min} & = & 2.49\times 10^7 \left(\frac{Ro}{Ro_\odot}\right)^{-0.03}\left(\frac{M_\star}{M_\odot}\right)^{0.83} {\rm cm}^{-3} \\
    T_c^{\rm min} & = & 1.5 \left(\frac{Ro}{Ro_\odot}\right)^{-0.04}\left(\frac{M_\star}{M_\odot}\right)^{0.05} {\rm MK}
\end{array} \right.\, , \\
&\left\{\begin{array}{ccc}
    n_c^{\rm max} & = & 4.63\times 10^8 \left(\frac{Ro}{Ro_\odot}\right)^{-1.07}\left(\frac{M_\star}{M_\odot}\right)^{1.97} {\rm cm}^{-3} \\
    T_c^{\rm max} & = & 1.5 \left(\frac{Ro}{Ro_\odot}\right)^{-0.17}\left(\frac{M_\star}{M_\odot}\right)^{0.19} {\rm MK}
\end{array} \right. \, .
\end{align}
Their values are given in Table \ref{ta:Windprop}. These values are the most extreme ones allowed by the modelling developed in \citet{Ahuir2020}. Note that they still do not reach the high values of $T_c$ and $n_c$ considered in \citet{Kavanagh2019} (K19 in Table \ref{ta:Windprop}), and that they bracket the values predicted from the pioneering model of \citet{Holzwarth2007} (HJ07 in Table \ref{ta:Windprop}). For the sake of completeness, we decided here to consider four sets of $(n_c,T_c)$ values, including the extreme case of  \citet{Kavanagh2019}. Therefore, the present work covers the largest acceptable parameter space for $T_c$ and $n_c$. We note that in \citet{Kavanagh2019} a polytropic index of $\gamma=1.1$ was used, whereas in the remaining three models considered here we chose to use $\gamma=1.05$ as in \citep{Strugarek2019}. This difference nevertheless does not significantly change the conclusions of the work presented here. \moddAS{Finally, the polytropic approach used in this work is a crude approximation to the real physical mechanisms heating the corona. More realistic models taking self-consistently the heating of the corona by Alfv\'en waves should be ultimately used to model asterospheres \citep[\textit{e.g.}][]{Alvarado-Gomez2016b,Reville2020}. Such models require to set other parameters such as the Poynting flux going into the corona \citep[\textit{e.g.}][]{Hazra2021}. We note nonetheless that following the work of the \citet{Ahuir2020}, the range of polytropic models we consider here encompasses a large parameter space within which the mass-loss and angular momentum loss of an Alfv\'en-wave driven model would fall. The 3D structure of the corona could nevertheless be different from the one obtained with the polytropic approximation explored in this work \citep[\textit{e.g.}][]{Hazra2021}, and we leave its detailed study for future work.}

\begin{figure}
  \centering
  \includegraphics[width=\linewidth]{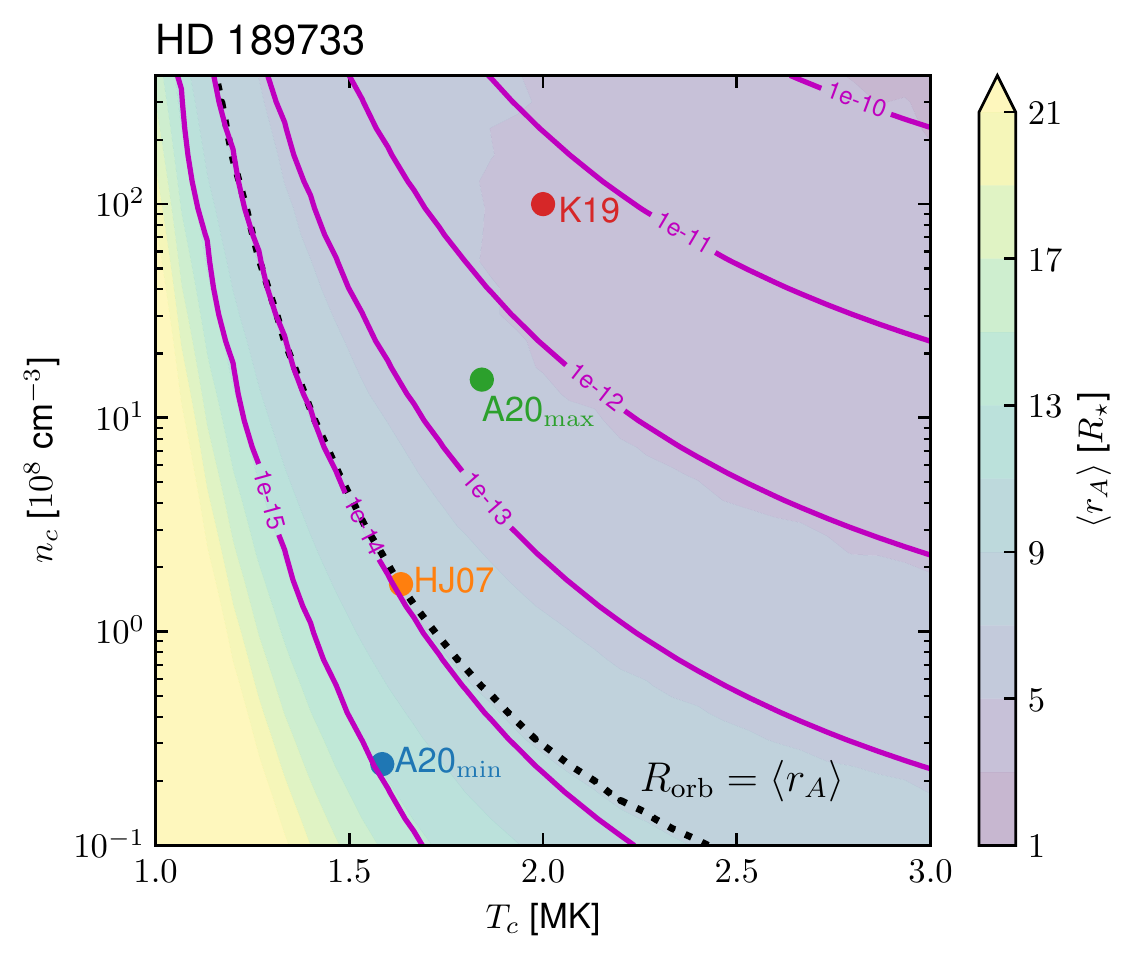}
  \caption{Average Alfv\'en surface and mass-loss predicted with the starAML \citep{Reville2015} package, based on the magnetic ZDI map of August 2013 for HD 189733. The results are shown as a function of the coronal density $n_c$ (y-axis) and of the coronal temperature $T_c$ (x-axis). The background colormap shows the average Alfv\'en surface in units of stellar radii, and the magenta contours label constant mass-loss rate in units of solar mass per year. The four modelling choices presented in this work (A20$_{\rm min}$, HJ07, A20$_{\rm max}$, K19, see text) are labelled by the colored dots. The average Alfv\'en radius corresponding the semi-major axis of HD 189733b is shown by the black dashed line.}
  \label{fig:WindProp}
\end{figure}

Before performing the 3D modelling with the PLUTO code, the global properties of the expected stellar wind can be estimated using the 1D open-source starAML code that computes a 1D Weber-Davis polytropic wind solution \citep{Reville2015a}\footnote{The source code is accessible \href{https://github.com/vreville/starAML}{here}}. Based on the values of $T_c$ and $n_c$, and on the magnetic map of August 2013, starAML can carry out a prediction of the mass loss rate of the wind model. We show the predicted mass-loss rate $\dot{M}_{\rm th}$ with this reduced model in the third column of Table \ref{ta:Windprop}. We see that the range of parameters considered here affect significantly the mass-loss rate, which varies by 3 orders of magnitude from model A20$_{\rm min}$ to model K19. The expected mass-loss rate variation as a function of $(T_c,n_c)$ is summarized in Fig. \ref{fig:WindProp}. Constant mass-loss rate in this diagram are shown by the purple contours, labelled by the mass loss value in units of $M_\odot$ yr$^{-1}$. The four modelling choices studied in this work are indicated by the colored-points (the same color-code will be used throughout the paper). The background color map shows the expected average Alfv\'en radius $\left\langle r_A \right\rangle$ based on the amplitude of the magnetic field from the ZDI map of August 2013. \moddAS{The starAML package computes the wind mass loss $\dot{M}_{\rm th}$ and the angular momentum loss $\dot{J}_{\rm th}$. The average Alfv\'en radius can be estimated with 
\begin{equation*}
    \left\langle r_A \right\rangle = \sqrt{\frac{P_{\rm rot}\dot{J}_{\rm th}}{2\pi \dot{M}_{\rm th}}}\, .
\end{equation*}}
It varies from a few stellar radii in the top right corner to more than 15 stellar radii in the lower left corner. The particular case where the semi-major axis of HD 189733b is equal to spherically symmetric averaged Alfv\'en radius $\left\langle r_A \right\rangle$ is shown by the black dotted line. Above this line, the planet is likely to spend the major part of its orbit outside the stellar wind Alfv\'en surface. Below the black dashed line, we expect it to be within the Alfv\'en surface most of the time. We remark that the A20$_{\rm min}$ and HJ07 models likely lead to a state where the planet is within the Alfv\'en surface, while the other two are in the opposite situation. Indeed, \citet{Kavanagh2019} found in their model that HD 189733b was likely always outside this surface. \moddAS{As we will see in what follows, the 3D Alfv\'en surface is asymmetric and therefore HD 189733b could actually go in and out of it along the orbit.}  

\begin{table}
	\centering
	\caption{Properties of the HD 189733 system}
	\label{ta:HD189733prop}
	\begin{tabular}{lrr} 
	\hline
	Parameter & Value & Reference \\
	\hline
  $T_{\rm eff}$ [K]   & 5050 $\pm$ 50   & \citet{Bouchy2005} \\ 
  $M_\star$ [$M_\odot$] & 0.92 $\pm$ 0.03  &  \citet{Bouchy2005}  \\ 
  $R_\star$ [$R_\odot$] & 0.76 $\pm$ 0.01 & \citet{Winn2007a}  \\ 
  $P_{\rm rot}$ [days]  & 11.94 $\pm$ 0.16 & \citet{Fares2010a}  \\
  \hline
  $M_p$ [$M_{\rm J}$]       & 1.13 $\pm$ 0.03 & \citet{Boisse2009}  \\
  $P_{\rm orb}$ [days]       & 2.2185733 $\pm$ 0.0000019 & \citet{Boisse2009} \\
  Semi-major axis [$R_\star$] & 8.86 $\pm$ 0.40 & \citet{Boisse2009} \\
  Semi-major axis [$R_p$] & 56.8 $\pm$ 2.00 & \citet{Boisse2009}\\
  Inclination angle $i$ [$^\circ$] & 85.78 $\pm$ 0.03 & \citet{Morello2014} \\
  \hline
	\end{tabular}
\end{table}


\begin{table}
	\centering
	\caption{Properties of the wind of HD 189733, derived here from the 1D starAML code.}
	\label{ta:Windprop}
	\begin{tabular}{lrrr} 
	\hline
	Model & $n_c$ [$10^8$ cm$^{-3}$] & $T_c$ [MK] &  $\dot{M}_{\rm th}$ [$10^{-14} M_\odot$ yr$^{-1}$] \\
	\hline
  A20$_{\rm min}$ &  0.24  & 1.59 &  0.15 \\
  HJ07  &  1.66  & 1.64 &  1.09 \\
  A20$_{\rm max}$ & 15.13  & 1.84 & 24.41 \\
  K19   & 99.84  & 2.00 &146.46 \\
  \hline
	\end{tabular}
\end{table}

\subsection{Magnetic field extrapolation techniques}
\label{sec:magExtrapol}

The second critical hypothesis behind the modelling of stellar coronae lies in the magnetic extrapolation technique. \moddAS{To initialize a stellar coronal model in 3D, a first extrapolation of the magnetic field in the computational domain is required. After this initialization step, the vector magnetic field is imposed only at the bottom boundary and the 3D domain evolves self-consistently under the MHD approximation.}  The ZDI methodology gives information on the vector magnetic field at the stellar surface. However, most models of stellar coronae rely only on the radial component of this field to initialize their 3D domain and to impose boundary conditions, loosing some observational constraints given by ZDI (\textit{e.g.} on the total field strength).  
A pioneering attempt was carried out by \citet{Jardine2013} with the inclusion of non-potential field at the initiation stage of their wind model. This non-potential component was then not retained in their boundaries, \moddAS{as they only maintained the radial component of the magnetic field there during the MHD simulations}. They concluded that these components have no significant impact on the modelling of the corona. We revisit here this initial work to assess the influence of the component selection on the driving and sculpting of the stellar corona.

\begin{figure*}
  \centering
 \includegraphics[width=0.99\linewidth]{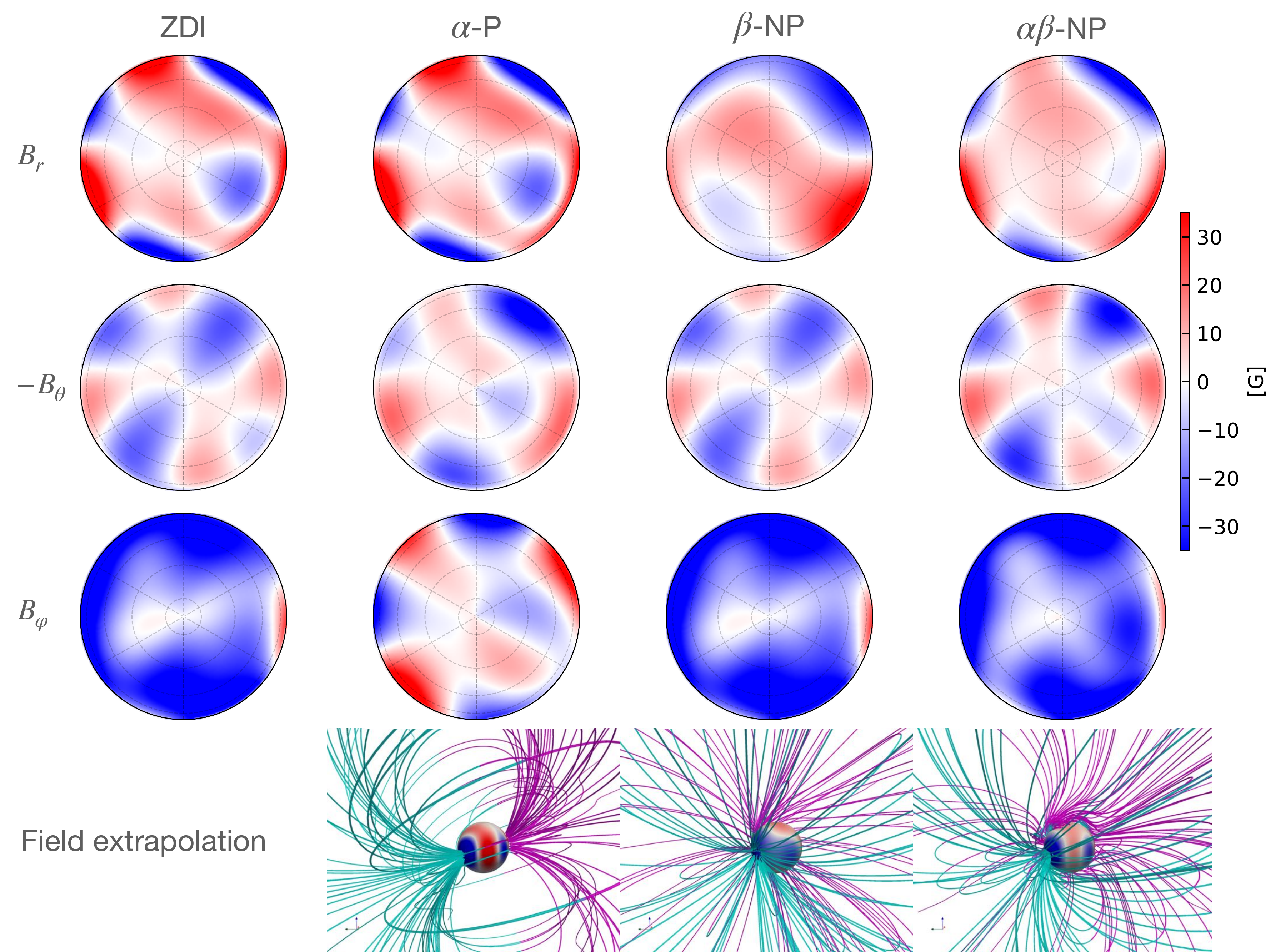}
  \caption{Magnetic maps of HD 189733. The left column shows the north hemisphere of the magnetic map derived with ZDI on the observation of August 2013 by \citet{Fares2017a}. The concentric dashed black circles correspond to latitudes 80$^\circ$, 60$^\circ$, 40$^\circ$, and 20$^\circ$. From top to bottom, the magnetic components in spherical coordinates are shown on a colored scale from -35 G (blue) to +35 G (red). Note that the second row shows the opposite of the co-latitudinal component of ${\bf B}$, for the sake of easing the comparison with the maps published in the literature (\textit{e.g.} \citealt{Fares2017a}). The second, third and fourth columns show the surface field reconstructed with the $\alpha$-P (\S\ref{sec:potent-extr-1}), $\beta$-NP (\S\ref{sec:non-potent-extr}) and $\alpha\beta$-NP (\S\ref{sec:non-potent-extr-mixt}) extrapolation methods. The last row illustrates the magnetic field in the computational domain at the initialization of the MHD runs. The magnetic field lines are colored in magenta ($B_r>0$) and cyan ($B_r<0$), and the surface radial magnetic field is shown.}
  \label{fig:MagMap}
\end{figure*}

We consider three different magnetic field extrapolation techniques, with mathematical details given in Appendix \ref{sec:Extrapol}. The first \moddAS{intialization} technique is dubbed $\alpha$-P and is a standard potential field with source surface extrapolation technique \citep[\textit{e.g.}][]{Altschuler1969,Schrijver2003c}, based on the radial component of the magnetic field only. This technique assumes that the radial field at the stellar surface is known, and that the field becomes purely radial outside an outer source surface $R_{\rm ss}$. In this work, we consider $R_{\rm ss}=15 R_\star$. 
Note that the exact value of $R_{\rm ss}$ affects the initial extrapolation in the whole computational domain and is used to set the fixed magnetic field at the bottom boundary. The wind then develops and changes self-consistently the magnetic connectivity in the corona. We have verified that the exact value of $R_{\rm ss}$ does not change significantly the steady-state MHD solution, since the opening of the field lines is dictated mainly by the wind itself rather than initial source-surface. The second \moddAS{initial} extrapolation, dubbed $\beta$-NP, is a non-potential extrapolation technique that is driven by the horizontal components of the stellar magnetic field ($B_\theta$ and $B_\varphi$ in spherical coordinates). 
Finally, the third \moddAS{initial} extrapolation dubbed $\alpha\beta$-NP combines the two previous methodologies to obtain a solution approaching the three components of the ZDI maps while retaining a tractable analytical formulation. The three methods are illustrated in Figure \ref{fig:MagMap}. The leftmost column shows the three components of the magnetic field obtained through ZDI for August 2013, as seen from the rotational north pole. The three extrapolation techniques are then illustrated in the three last columns with the same color map varying from -35 to 35 G. The extrapolated field components are shown at the base of the domain in the first three rows, and a 3D field line representation of the extrapolated field is shown in the last row. We remark that by design, the $\alpha$-P extrapolation exactly matches the radial component of the ZDI field, but produces a surface horizontal field that significantly differs from the ZDI map. 
Conversely, the $\beta$-NP extrapolation matches the horizontal field, but produces a different surface radial field. 
Finally, the $\alpha\beta$-NP does not match exactly any of the three components, but does match their overall variations and can therefore be considered as the closest extrapolation to the 3D vector magnetic field. \modAS{We report the energetics of the surface field in Table \ref{ta:Emodels}. The energies are calculated by performing the integral of $B^2$ over the spherical surface divided by $4\pi$, following equation \ref{eq:integralB}. The rms magnetic field $\left\langle B \right\rangle_{\rm rms}$ (second column) is obtained by taking the square root of the energy. We see that the rms field strength varies slightly, from 33.9 to 40.8 G, depending on the field reconstruction technique. The total energy $E_{\rm tot}$ characterizes the full magnetic field (third column), the toroidal energy $E_{\rm toro}$ (third column) uses only the components deriving from the $\gamma^m_l$ spherical harmonics coefficients (see Appendix \ref{sec:Extrapol}), and the axisymmetric energy $E_{m=0}$ (fourth column) uses only the $m=0$ poloidal field. The original ZDI map has 50\% of its energy within the toroidal components (see also \citealt{Fares2017a}) and only 0.9\% of its energy in the poloidal axisymmetric field. The $\alpha$-P model possesses no toroidal field \modAS{at initialisation}, and generates a surface field that is 30\% less energetic than the original ZDI field (fifth column). Model $\beta$-NP possesses the strongest relative toroidal field, and model $\alpha\beta$-NP approaches best the ZDI map while being only 17\% less energetic. In all models, the axisymmetric poloidal field remains negligible. As a consequence of these differences, we see in the last row of Fig. \ref{fig:MagMap} that the three extrapolations lead to a very different magnetic connectivity in the environment of the star. This will have a significant impact on the predicted star-planet magnetic interaction signal, as we will see in the next sections.}

\begin{table}
	\centering
	\caption{Energetics of the magnetic field on the stellar surface}
	\label{ta:Emodels}
	\begin{tabular}{lrrrr}
	\hline
	 & $\left\langle B \right\rangle_{\rm rms}$ [G] & $E_{\rm tor}/E_{\rm tot}$ & $E_{m=0}/E_{\rm tot}$ & $E_{\rm tot}/E_{\rm tot,ZDI}$  \\
	\hline
  ZDI               &  40.8  & 0.50 &  0.009 & 1.0  \\
  $\alpha$-P        &  34.0  & 0.00 &  0.009 & 0.70 \\
  $\beta$-NP        &  35.8  & 0.65 &  0.017 & 0.77 \\
  $\alpha\beta$-NP  &  37.3  & 0.60 &  0.009 & 0.83 \\
  \hline
	\end{tabular}
\end{table}

We recall that the extrapolation techniques are used at initialization in our model. The magnetic field and the stellar wind are then left free to evolve self-consistently in the computational domain, \moddAS{and the three components of} the magnetic field are maintained to their initial values at the stellar boundary. We now turn to the effect of the different wind modelling hypothesis on the predicted corona of HD 189733.

\section{Modelling of HD 189733 on August 2013}
\label{sec:multiEpoch}

We summarize the modelled properties of the wind of HD 189733 in Table \ref{tab:WindModels}. We report the mass-loss rate of each model, the average radius of the Alfv\'en surface on the orbital plane $\left\langle r_A \right\rangle_{\rm orb}$, and the open flux $\Phi_{\rm open}$ in the corona of the model. We see that the mass loss rate is coherent with the predicted value from Fig. \ref{fig:WindProp}, and does not vary significantly with the extrapolation method. This is expected from polytropic wind as the wind driving originates mostly from the assumed coronal temperature $T_c$ \citep{Reville2016a}, and the mass loss rate is therefore only mildly affected by the magnetic field itself. 

\begin{figure*}
  \centering
  \includegraphics[width=0.32\linewidth]{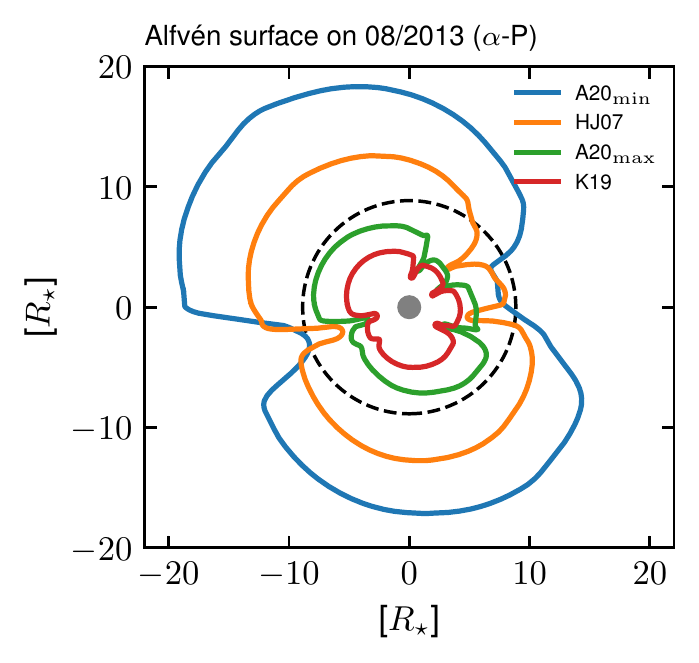}
  \includegraphics[width=0.32\linewidth]{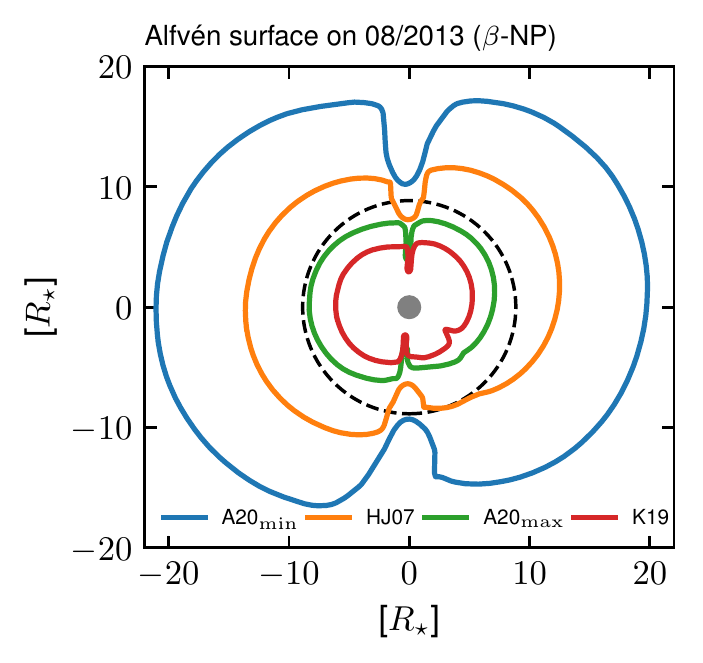}
  \includegraphics[width=0.32\linewidth]{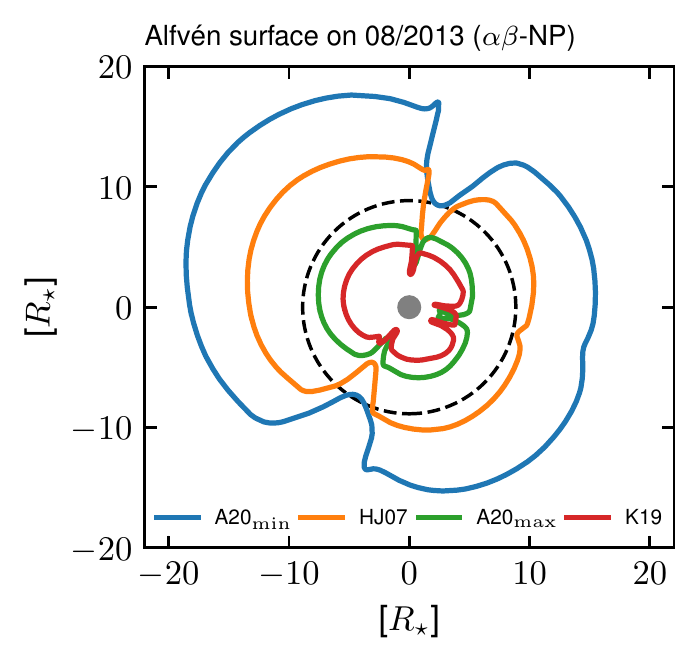}
  \caption{Contours of the Alfv\'en surface on the orbital plane, defined as $M_a=1$ (see text). Each model (A20$_{\rm min}$, HJ07, A20$_{\rm max}$, K19) is represented by a different color (blue, orange, green, red). From left to right, models $\alpha$-P, $\beta$-NP and $\alpha\beta$-NP are shown. The orbit of HD 189733b is shown in each panel by the dashed black line.}
  \label{fig:CompAlfSurf}
\end{figure*}

We define here the Alfv\'en surface based on the relative motion of the stellar wind in a frame where the orbiting planet is at rest. This average Alfv\'en surface is reported in column 4 of Table \ref{tab:WindModels} (with maximum and minimum values as superscripts and subscripts), and is plotted on the orbital plane in Figure \ref{fig:CompAlfSurf} for all models. From left to right, the three magnetic extrapolation techniques are shown and in each panel the four models A20$_{\rm min}$ (blue), HJ07 (orange), A20$_{\rm max}$ (green) and $K19$ (red) are shown. The size of the Alfv\'en surface decreases with $T_c$ and $n_c$. In each panel, the circular orbit of HD 189733b is shown by a black dashed line. The modelling choices for the corona and wind of HD 189733 have a strong impact on the position of the Alfv\'en surface relative to the orbit. \moddAS{Models HJ07 and A20$_{\rm min}$ predict that the planet is almost always inside the Alfv\'en surface, and in the case of model A20$_{\rm min}$, $\beta$-NP it is never outside it. Conversely, the K19 and $A20_{\rm max}$ models predict that HD 189733b is always outside the Alfv\'en surface, \textit{i.e.} it orbits in the external super-alfv\'enic part of the stellar atmosphere. In the case of model $A20_{\rm max}$, the orbit of the planet sometimes almost reaches the boundary of the Alfv\'en surface (see \textit{e.g.} the middle panel). If that is indeed the case, no magnetic connection is possible between the star and the planet in these cases, and this would imply that the signal detected by \citet{Cauley2018} cannot originate from SPMI. 
 We recall that the A20$_{\rm min}$ and A20$_{\rm max}$ models are based on the study of \citet{Ahuir2020} and bracket the most probable parameter space for modelling the wind of HD 189733 with a polytropic approach. Based on this study, model A20$_{\rm max}$ corresponds to the case with the largest mass loss and the smallest Alfv\'en surface. Therefore, since in the cases A20$_{\rm max}$ the planet almost reaches the borders of the Alfv\'en surface at some orbital phases, we conclude that it is likely that HD 189733\,b spends at least part of its orbit within the Alfv\'en surface of its host star wind, and that SPMIs affecting the star could therefore be in action in this system.}

 The Alfv\'en surface varies significantly in the three panels of Figure \ref{fig:CompAlfSurf}. In each panel, the surfaces have a similar shape but change in size. From one panel to the other, we see that the shape of the Alfv\'en surface changes due to a different magnetic topology induced by the extrapolation choice \moddAS{and can rotate by an angle close to 90$^\circ$}. Such changes do not affect the fundamental SPMI, but change its temporal variability along the planetary orbit.

\begin{table}
	\centering
	\caption{Properties of the 3D MHD models of the wind of HD 189733. From left to right, the column are the model name, the type of magnetic field extrapolation (see \S\ref{sec:magExtrapol}), the mass loss rate of the wind, the average Alfv\'en radius on the orbital plane, and the open flux in the stellar wind.}
	\label{tab:WindModels}
	\begin{tabular}{lrrrr} 
	\hline
	Model & Magnetic & $\dot{M}$ &  $\left\langle r_A \right\rangle_{\rm orb}$ & $\Psi_{\rm open}$ \\
	 & extrapolation & [$10^{-14} M_\odot$ yr$^{-1}$] &  $^{\rm max}_{\rm min}$[$R_\star$] & [$10^{22}$ Mx] \\
	\hline
  \multirow{3}{*}{A20$_{\rm min}$} & $\alpha$-P & 0.15 & $15.1_{7.4}^{20.6}$ & 0.7 \\
                                   & $\beta$-NP & 0.15 & $16.6_{9.3}^{21.1}$ & 0.6 \\
                             & $\alpha\beta$-NP & 0.18 & $15.8_{8.5}^{19.8}$ & 0.6 \\\hline 
  \multirow{3}{*}{HJ07}            & $\alpha$-P & 1.1 & $10.8_{4.5}^{14.1}$ & 1.1 \\ 
                                   & $\beta$-NP & 2.6 & $11.0_{6.4}^{13.6}$ & 1.1 \\
                             & $\alpha\beta$-NP & 1.5 & $10.8_{5.5}^{14.2}$ & 1.2 \\\hline
  \multirow{3}{*}{A20$_{\rm max}$} & $\alpha$-P & 24.3 & $5.9_{2.6}^{8.1}$ & 2.2 \\
                                   & $\beta$-NP & 30.8 & $6.8_{3.3}^{8.4}$ & 3.0 \\
                             & $\alpha\beta$-NP & 29.7 & $5.7_{2.5}^{7.9}$ & 2.4 \\\hline 
  \multirow{3}{*}{K19}            & $\alpha$-P & 146.0 & $4.1_{2.1}^{5.5}$ & 3.0  \\
                                  & $\beta$-NP & 161.0 & $5.0_{2.3}^{6.1}$ & 4.5 \\
                            & $\alpha\beta$-NP & 147.0 & $4.2_{2.1}^{5.7}$ & 3.4 \\
  \hline
	\end{tabular}
\end{table}

\begin{figure*}
  \centering
  \includegraphics[width=\linewidth]{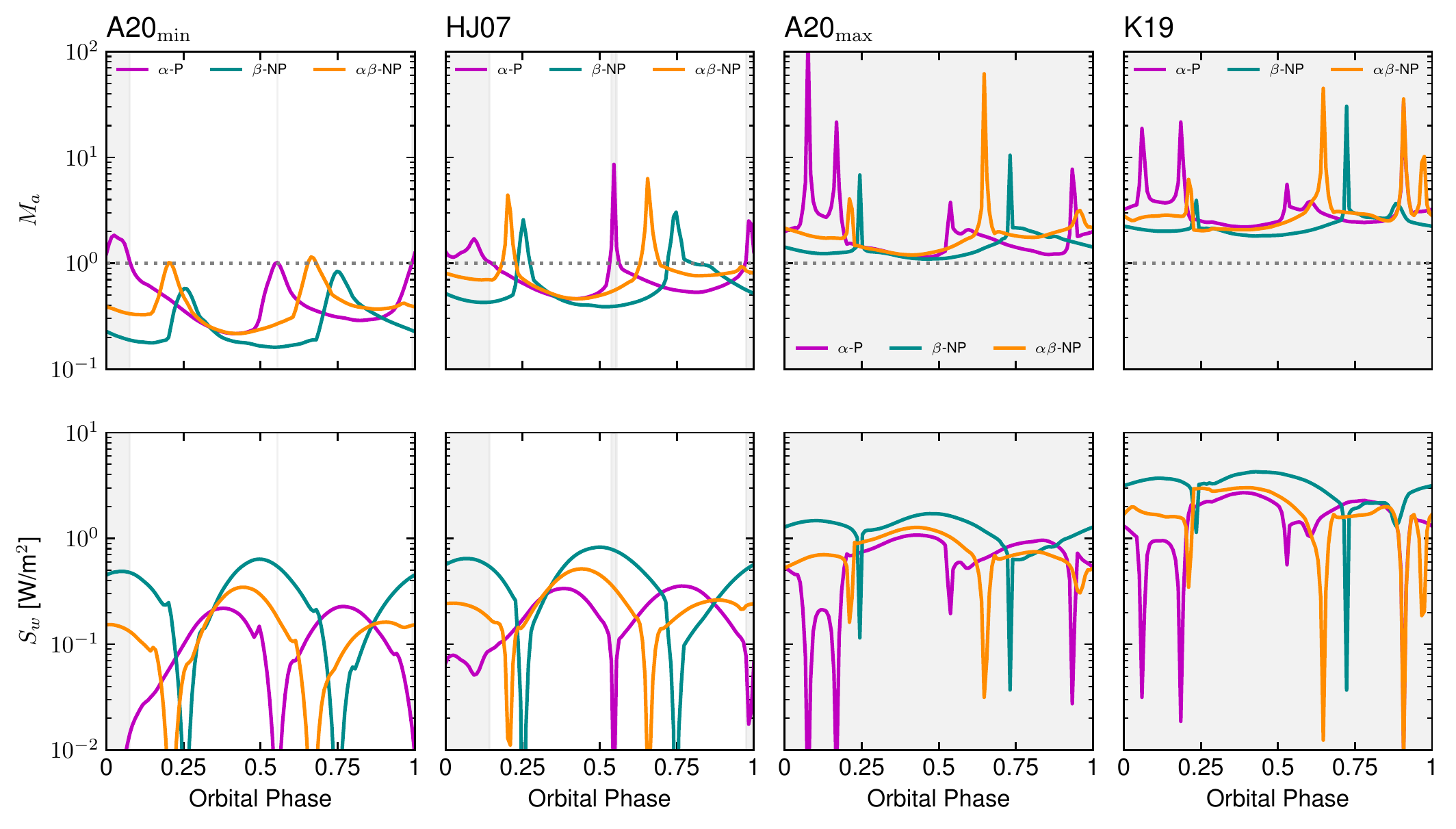}
  \caption{Properties of the star-planet magnetic interactions along the orbit of HD 189733. The first row shows the Alfv\'enic Mach number $M_a$, and the second row the Poynting flux amplitude $S_w$ in units of W/m$^2$. Each column correspond to one of the four modelling choices (from left to right A20$_{\rm min}$, HJ07, A20$_{\rm max}$, and K19). In each panel, the three magnetic extrapolation are shown in magenta ($\alpha$-P), green-blue ($\beta$-P) and orange ($\alpha\beta$-NP). The greyed areas correspond to phases where $M_a>1$ in case $\alpha-P$. Here, the orbital phase is arbitrarily set to 0 when the planet is at the zero longitude of the magnetic maps shown in Fig. \ref{fig:MagMap}. The real planet phase with respect to the rotating magnetic map is taken into account only in the comparison with the observational data in Fig. \ref{fig:SPMIsignal}.}
  \label{fig:PropAlongOrbit}
\end{figure*}

We show in Figure \ref{fig:PropAlongOrbit} the properties of the SPMI along the orbital path for each model. The upper panel shows the Alfv\'enic Mach number $M_a=\tilde{v}/v_a$ where $\tilde{v}$ is the velocity of the interplanetary medium in the frame orbiting with the planet and $v_a=B/\sqrt{4\pi\rho}$ is the local Alfv\'en speed. As expected from the orbital plane shown in Fig. \ref{fig:CompAlfSurf}, we recover that the planet is predicted to be inside or outside the Alfv\'en surface depending on its orbital phase. For instance, in model HJ07 (second column) the planet is in a sub-alfv\'enic interaction regime ($M_a<1$) most of the time, and in a super-alfv\'enic interactions regime ($M_a>1$) near $\phi_{\rm orb}=0.15$ and $\phi_{\rm orb}=0.51$ for model $\alpha$-P (magenta line). The other models ($\beta$-NP and $\alpha\beta$-NP) present a similar profile, shifted by a few tenths of planetary orbit. Conversely, we recover also that model K19 predicts a planet always in a super-alfv\'enic interaction regime with $M_a>1$ for all orbital phases.  

The lower panel of Fig. \ref{fig:PropAlongOrbit} shows the Poynting flux density in the stellar wind accessible to SPMI as a function of the orbital phase. This Poynting flux density is denoted $S_w$ and is defined as 
\begin{equation}
    \label{eq:Sw}
    S_w = -\frac{1}{\mu_0} \left[\left({\bf v}\times {\bf B}\right) \times {\bf B}\right] \cdot \frac{\tilde{{\bf v}}}{|\tilde{{\bf v}}|}\, .
 \end{equation}
The Poynting flux density varies from a few tenths of W/m$^2$ to a few W/m$^2$ in the 12 models considered here. It can vary significantly along the orbit of the planet, due to the large variations expected in the amplitude of the wind magnetic field there \citep[\textit{e.g.}][]{Fares2010a}. For instance, in model HJ07 ($\alpha$-P) in magenta in the second column, the Poynting flux density varies from 0.003 to 0.35 W/m$^2$ along one orbit. Finally, the SPMI is therefore fed with a highly varying Poynting flux along the orbit. We consequently expect that any SPMI signal should present significant variability with an orbit solely based on this variation. \moddAS{We also note that in the super-alf\'venic cases (last and second-to-last columns), the SPMI still intersects this Poynting flux density but the SPMI channels energy towards the interplanetary medium. This flux could still be meaningful, \textit{e.g.} for the SPMI in the context of radio emission within planetary magnetospheres forced by its interaction with the stellar wind.}

In our approach, $\alpha$-P models generally predict the smallest Poynting flux density levels, and $\beta$-NP models predict the largest levels. This correlates with the relative toroidal energy of the reconstructed surface magnetic field (first column in Table \ref{ta:Emodels}) used to drive the coronal model. The Poynting flux density varies at most by a factor of 2 when changing the extrapolation method, therefore we consider that the estimated levels of $S_w$ shown in Fig. \ref{fig:PropAlongOrbit} are robust. Note that models with higher mass loss rates (from left to right) generally predict larger maximum levels of $S_w$.

The large variations of $M_a$ and $S_w$ along the orbital path leads to the conclusion that any SPMI signal should also embed such variability. We now turn to the estimate of the amplitude and relative phasing of the SPMI signals expected from our set of simulations that exhibit sub-alfv\'enic orbits.

\section{Expected star-planet magnetic interactions}
\label{sec:ExpSPMI}

\subsection{Amplitude of star-planet magnetic interactions}
\label{sec:SPMImodels}

In the recent literature, several flavors of star-planet magnetic interactions have been considered to estimate the power associated with them. 

On one side, the Alfv\'en-wings (hereafter AW) SPMI was proposed for compact exosystems based on the parallel drawn with such interactions in planet-satellite interactions within the solar system. Following the analytical development of \citet{Saur2013} that refined the original developments of \citet{Zarka2001,Zarka2007}, an estimate of the maximal power channeled by SPMI towards the central star (provided the planet orbits within the Alfv\'en surface) can be written as
\begin{equation}
    \label{eq:PowerS13}
    \mathcal{P_{\rm AW}} = 2\pi R_P^2 S_w \left( 3 \zeta^{-2/3} M_a \right)\, ,
\end{equation}
where $\zeta=B_w/B_P$ is the ratio of the interplanetary magnetic field at the planet orbit $B_w$ to the assumed planetary magnetic field $B_P$ at the surface of the planet. The factor $3$ originates from the fact that the effective obstacle is actually the planet and its magnetosphere, and therefore in the most efficient magnetic topology the obstacle increases by a factor 3 (see \citealt{Saur2013} for more details). We also have dropped an efficiency factor ($\bar{\alpha}$ in \citealt{Saur2013}) to provide an estimate of the maximal power involved in this model of SPMIs.

On the other side, \citet{Lanza2013a} proposed a different interpretation of SPMI whereby the accessible power is directly tapped from the planetary field itself. In this scenario, the planetary field reconnecting with the ambient stellar wind field is stretched by the orbital motion of the planet and ultimately breaks by means of reconnection. We dubbed this scenario as \textit{stretch-and-break} (SB) in what follows. The SB power can be estimated as

\begin{equation}
    \label{eq:PowerL13}
    \mathcal{P_{\rm SB}} = 2\pi R_P^2 S_w \left(\zeta^{-2} f_{\rm AP}\right)\, ,
\end{equation}
where $f_{\rm AP}$ is the area fraction of the planetary disk where magnetic field lines are connected to the ambient wind. In the optimal magnetic topology where the polar planetary field is oriented in the same direction as the ambient field, it can be estimated as 
\begin{equation}
    \label{eq:fAP}
    f_{\rm AP} = 1 - \left(1 - \frac{3\zeta^{1/3}}{2+\zeta} \right)^{1/2}\, .
\end{equation}

The ratio $\mathcal{P}_{\rm SB}/\mathcal{P}_{\rm AW}$ can reach a 100 to a 1000, depending on the Alfv\'enic Mach number $M_a$ and the magnetic field ratio $\zeta$. In an attempt to model such interactions from first principles in 3D,
\citet{Strugarek2016c} found in numerical MHD simulations that the Alfv\'en-wings SPMI was indeed in action. In this work, he did not find hints of the stretch-and-break scenario of \citet{Lanza2013a}. It is today nevertheless difficult to rule completely out the fact that numerical simulations at higher resolution may lead to the development of such a type of SPMI. For the sake of completeness, we therefore consider in what follows the two possibilities, and leave for future work the exploration of the likeliness of the SB scenario.  

\begin{figure}
  \centering
  \includegraphics[width=\linewidth]{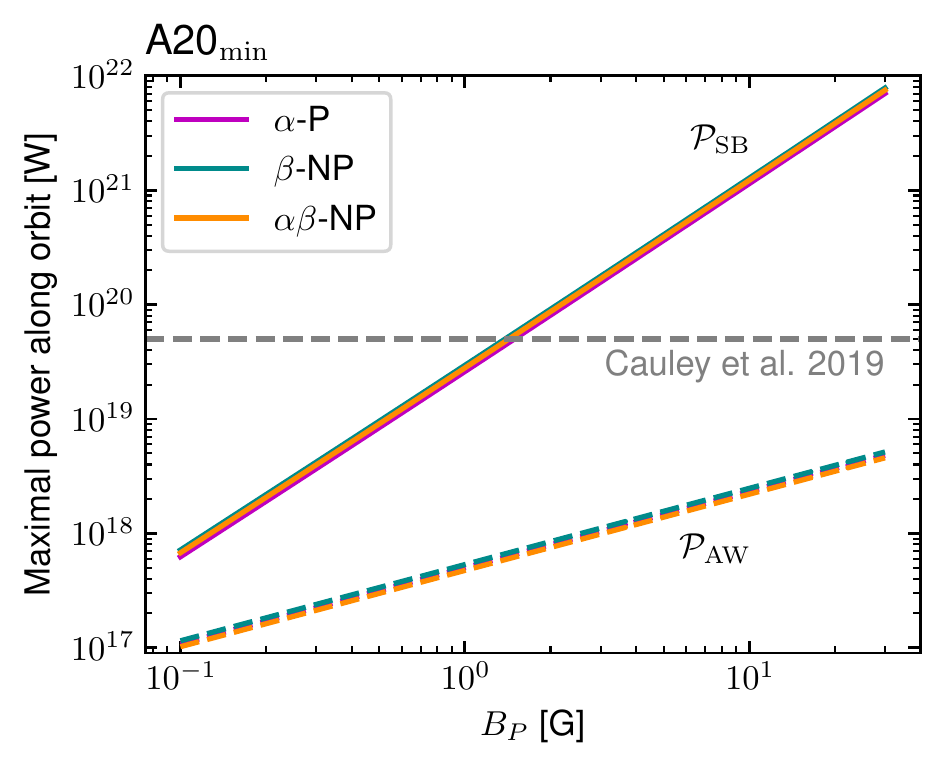}
  \includegraphics[width=\linewidth]{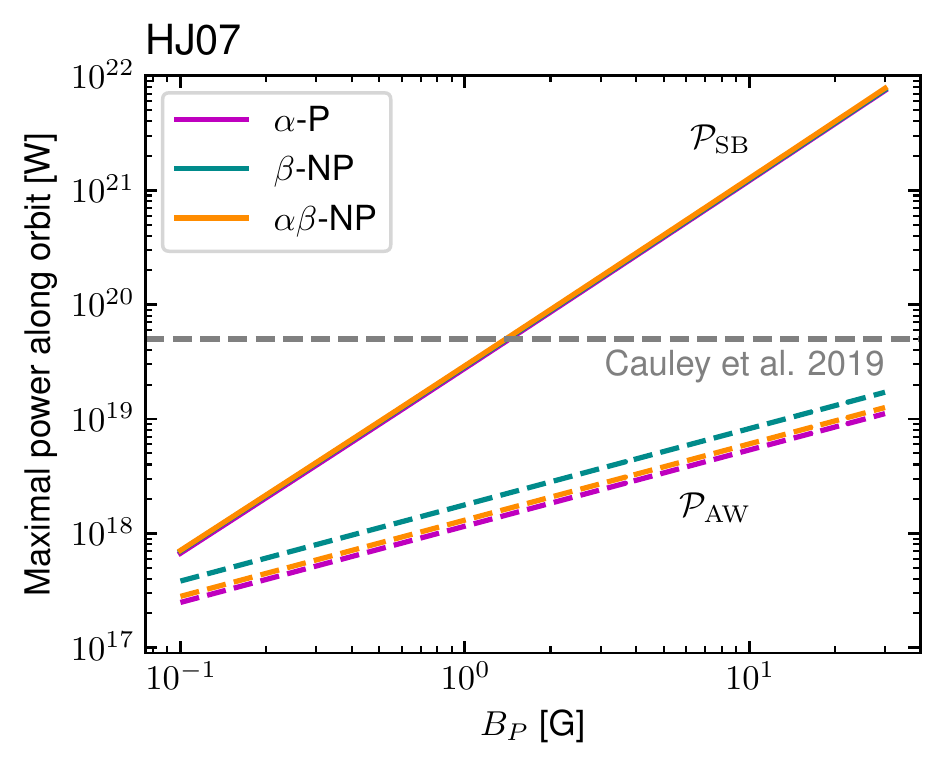}
  \caption{Maximal power (in W) available along the orbit as a function of the assumed planetary magnetic field $B_p$ (in G). The top panel corresponds to models A20$_{\rm min}$ and the bottom panel to models HJ07. The models are colored with respect to the magnetic extrapolation technique, as in Fig. \ref{fig:PropAlongOrbit}. The powers estimated with the stretch-and-break scenario ($\mathcal{P}_{\rm SB}$) are shown with a solid line, and the powers estimated with the Alfv\'en wings scenario ($\mathcal{P}_{\rm AW}$) with a dashed line. The maximum observational signal detected by \citet{Cauley2019} is shown by the grey horizontal dashed line.}
  \label{fig:PowerAlongOrbit}
\end{figure}

We show the maximum --along the orbit-- of $\mathcal{P}_{\rm AW}$ and $\mathcal{P}_{\rm SB}$ as a function of the assumed $B_P$ in Fig. \ref{fig:PowerAlongOrbit}. We consider planetary fields between 0.1 G and 30 G, following the estimated field proposed by \citet{Cauley2019}. The upper panel shows the maximal power reached in model A20$_{\rm min}$, and the lower panel the maximum power reached in model HJ07. The two other models are not shown here, as the planet is then outside the Alfv\'en surface for most of its orbit.
We find that $P_{\rm AW}$ can vary from $10^{17}$ W up to a maximum close to $10^{19}$ W. $P_{\rm SB}$ can reach much higher values up to about almost $10^{22}$ W. The Alfv\'en-wing scenario predicts powers that are more sensitive to wind model due to the additional $M_a$ dependency in Eq. \ref{eq:PowerS13}. Conversely, the amplitude of $\mathcal{P}_{\rm SB}$ predicted by \citet{Lanza2013a} varies even less with respect to the wind modelling choice, as seen when comparing the two panels and the different extrapolation techniques. 

\begin{figure}
  \centering
  \includegraphics[width=\linewidth]{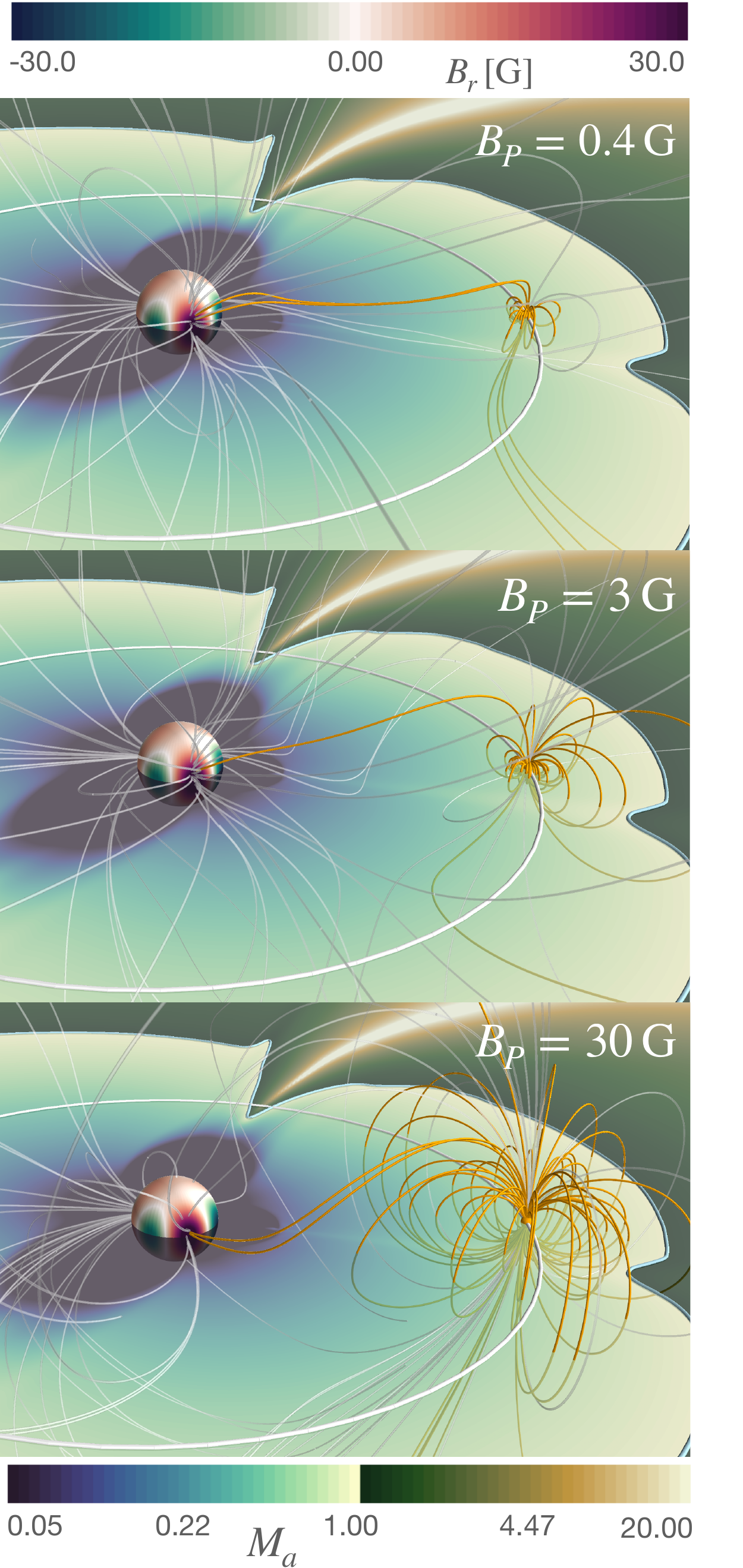}
  \caption{Three-dimensional rendering of the corona and wind of HD 189733 for model HJ07 ($\alpha\beta$-NP). The magnetic field lines are shown by the gray tubes. The Alfv\'enic Mach number $M_a$ is shown on the orbital plane in log scale, in bright blueish tones when $M_a < 1 $ and bright yellowish tones when $M_a>1$. The radial magnetic field at the bottom spherical boundary is shown in green (negative $B_r$) and red (positive $B_r$). The orbit of HD 189733 is illustrated by the white circle. The hypothetical magnetic field of HD 189733b, added here as a post-processing (\textit{i.e.} it is not included in the MHD simulation, and therefore does not take into account any pressure equilibrium and does not produce a tail), is shown by the orange tubes. From top to bottom, the planetary field is assumed to reach 0.4 G, 3 G and 30 G at the planetary surface.}
  \label{fig:DummyPlanets}
\end{figure}

\citet{Cauley2019} detected a signal that could originate from SPMI on HD 189733. They found that this signal corresponded to a power of about $5\,\times 10^{19}$ W (grey dashed line in Fig. \ref{fig:PowerAlongOrbit}). From rough scaling laws, they noted that $\mathcal{P}_{\rm SB}$ was \textit{a priori} the only SPMI model predicting strong enough interactions to explain such large powers. In this study we reach essentially a similar conclusion. 
Indeed, we recall here that these estimates are the maximum power available from SPMI. Only part of that power would be transferred to the CaII emission lines. We do not know today which conversion factor should be applied in the context of SPMI. Therefore, only the mechanism proposed in \citet{Lanza2013a} is likely predicting enough power to explain the observed signal with SPMIs, with $B_P \gtrsim 1 G$. The very large values of magnetic field reported by \citet{Cauley2019} account for a safe conversion factor, whereby a few tenths of a percent of the total SPMI power is converted into an observable CaII flux. Such large magnetic field must then be put in the context of the conditions at the planetary orbit. In Fig. \ref{fig:DummyPlanets} we have added a virtual planetary magnetosphere within our stellar wind model. Note that this magnetosphere, represented by the orange field lines, was not evolved self-consistently within the MHD wind solution. It must therefore be considered here only as an illustration, \moddAS{as we simply added a dipolar field to the simulation results at the planet location in post-processing. It nevertheless gives a rough idea of the size of the magnetosphere, because the pressure at the planetary orbit is dominated by the magnetic pressure of the wind which accounts for about 80\% of the total pressure there.} We show the size of the magnetosphere of the planet when considering a planetary field of 0.4, 3, and 30 G from top to bottom. We see that in the latter case, the magnetosphere fills up almost to the stellar surface. This extreme situation would require dedicated modelling where a self-consistent planet is embedded in the stellar wind (see \textit{e.g.} \citealt{Cohen2009b,Strugarek2015}). \refAS{Indeed, in this situation the magnetosphere can have subparts inside and outside the Alfv\'en surface, and the location of the footpoints of the interaction can vary significantly compared to the compact magnetosphere case shown in the top panel.} This aspect is out of the scope of the present study and will be addressed in a future work. 

It is nevertheless puzzling to understand why \citet{Cauley2018} detected a SPMI signal in August 2013, and not at other observational epochs. One possibility would be that the planet was within the Alfv\'en surface of the stellar wind only at that epoch. This is nevertheless not very likely, because the amplitude of the surface magnetic field of HD 189773 did not change significantly between the epochs studied in  \citet{Cauley2018} (see \citealt{Fares2010a}), and therefore the size of the Alfv\'en surface size likely did not change significantly either. Another source of difficulty to detect SPMI in stellar activity tracers is that their signal originates from energy deposition in the stellar chromosphere. The phase of such signal is a complex convolution between orbital motion, stellar rotation, and the magnetic topology in the stellar atmosphere that controls the travel-time of the Alfv\'en waves and particles carrying the SPMI energy from the vicinity of the planet down to the stellar chromosphere. It is therefore possible that at other epochs the SPMI signal exists but is not well correlated with the orbital period. We now turn to studying this complex interplay thanks to the 3D models presented in this work.  


\subsection{Phase of star-planet magnetic interactions}
\label{sec:PhaseSPMI}

\begin{figure*}
  \centering
    \includegraphics[width=\linewidth]{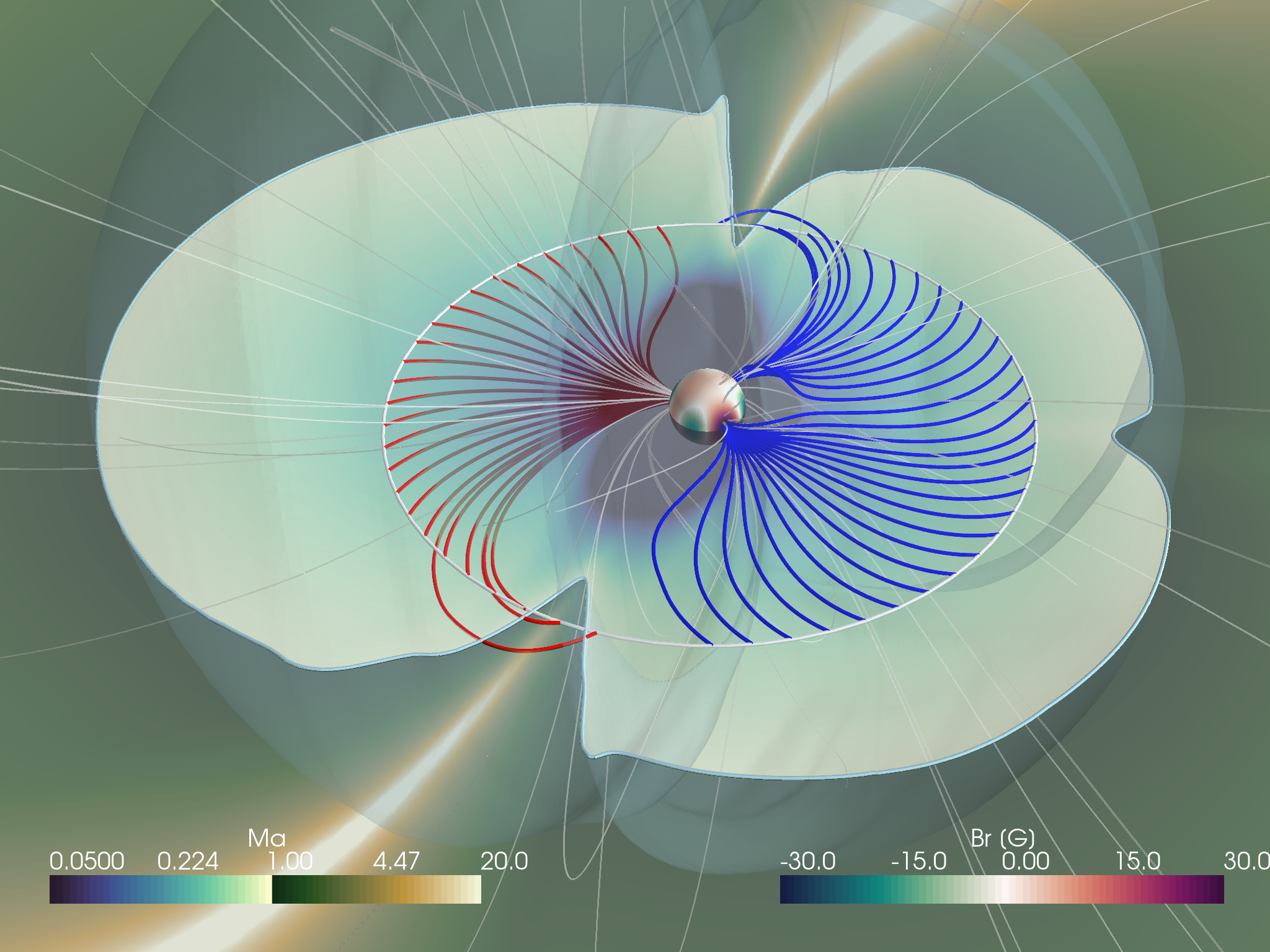}
  \caption{Three-dimensional rendering of the corona and wind of HD 189733 for model HJ07 ($\alpha\beta$-NP). The layout is the same in Fig. \ref{fig:DummyPlanets}. The Alfv\'en surface is shown by the transparent blue volume, and the light blue tube is its imprint on the orbital plane. The Alfv\'enic Mach number $M_a$ is shown on the orbital plane in log scale, in bright blueish tones when $M_a < 1 $ and bright yellowish tones when $M_a>1$. The magnetic field lines in the stellar corona and wind are illustrated with transparent gray tubes. The radial magnetic field at the bottom spherical boundary is shown in green (negative $B_r$) and red (positive $B_r$). The magnetic connectivity between the orbital path and the stellar surface is shown by the red (${\bf c}_a^-$) and blue (${\bf c}_a^+$) lines (see text).}
  \label{fig:3Dexample}
\end{figure*}

The power associated with SPMI is channeled from the planet vicinity towards the star along the Elsasser characteristics \citep{Strugarek2018a} defined as
\begin{equation}
    {\bf c}_a^{\pm} = {\bf v} \pm {\bf v}_a\, .
    \label{eq:Elsasser}
\end{equation}
These can be estimated based on the 3D structure of the model, and are illustrated for case HJ07 ($\alpha\beta$-NP) in Fig. \ref{fig:3Dexample}. The \moddAS{streamlines of} the Elsasser characteristics from the planet orbit (white circle) towards the star are shown in red (${\bf c}_a^-$) and blue (${\bf c}_a^+$). At each orbital phase, the two characteristics exist, but here we only show the ones that connect back to the star. The Alfv\'en surface of the stellar wind is shown by the transparent blue volume, and its imprint on the orbital plane is highlighted by a light blue tube. We immediately remark that in this case, the SPMI connects back to the star only on three different locations in Fig. \ref{fig:3Dexample}. This means that any SPMI signal imprinted on the stellar disk is visible only when at least one of these locations is visible for the observer. These locations correspond to the main magnetic poles of the star, and it implies that any induced hot spot will be differently phased along the orbit with respect to the planet location.

In order to be more quantitative, we make use of the 3D wind model to predict the observable phases of SPMI. The procedure involves the following steps
\begin{itemize}
    \item We use ephemerids of the planet to position the planet along its orbit relatively to the rotational phase of the star. Here we use the detailed ephemerids of HD 189733 published in \citet{Fares2017a}. We have tested other ephemerids (\textit{e.g.} \citealt{Agol2010,Hrudkova2010,Kokori2021}) and found no significant differences in our results for August 2013. 
    \item The SPMI signal originates from the vicinity of the planet in the two SPMI scenarios considered here (see \S \ref{sec:SPMImodels}). It is then transported towards the star along the Elsasser characteristics $c_a^{\pm}$ (see Eq. \ref{eq:Elsasser}). For the sake of completeness, we have also tried to transport the SPMI signal along the magnetic field lines rather than the Elsasser characteristic. Since the planet is close to its host, this makes negligible differences, and we therefore use the Elsasser characteristics in what follows.
    \item In the AW and SB scenarios, the energy is transported towards the star at approximately the local Alfv\'en speed. The Alfv\'en travel time from the orbit of the planet down to the star therefore varies along the orbit. For instance, in model HJ07 ($\alpha\beta$-NP) the travel time can vary between 1 and 35 hours. At maximum, this corresponds to 67 \% of the orbital period. This means that at these orbital phases, the energy available for SPMI reaches its final destination in the stellar chromosphere more than half an orbit after its trigger. Such delays are important and must therefore be accounted for when estimating the phase of SPMI signals. \moddAS{For the sake of completeness, we have also assessed the SPMI signal assuming an instantaneous information transport, which would be realized if the SPMI signal originated from accelerated electrons in the vicinity of the planet \citep[see \textit{e.g.}][]{Saur2018}. This is discussed in Appendix \ref{sec:Propagation sensivity}, and we found that the SPMI signal is modified only at some specific times by this assumption, and that the overall properties of the signal remain the same.}
    \item Stellar rotation must be taken into consideration when producing synthetic SPMI signal. We consider in this work that HD 189733 rotates in 12 days \citep{Fares2010a}, \moddAS{and that the ZDI map is representative of the magnetic topology of HD 189733 during the reference observations from \citet{Cauley2018} that were taken over about 18 days}. As the star rotates, we follow the impact location of the SPMI which can be within the visible stellar disk or behind the star, making the predicted SPMI signal visibility change (see \textit{e.g.} \citealt{See2015a}). \modAS{We note that stellar rotation is the most influential parameter on the temporal signature of the SPMI signal. We discuss the sensitivity of the signal with $P_{\rm rot}$ in Appendix \ref{sec:Prot_sensitivity} and consider here the canonical value of 12 days.} \moddAS{More advanced modelling shall include differential rotation as well, which is ignored for the time being in this work.}
    \item We take into account the inclination of the stellar axis of rotation with respect to the viewing angle from Earth. \citet{Fares2017a} estimate that the rotation axis of HD 189733 is almost perpendicular to the line of sight, with an inclination of approximately 5 degrees. We take into account this inclination in what follows.  
\end{itemize} 

\begin{figure*}
  \centering
    \includegraphics[width=\linewidth]{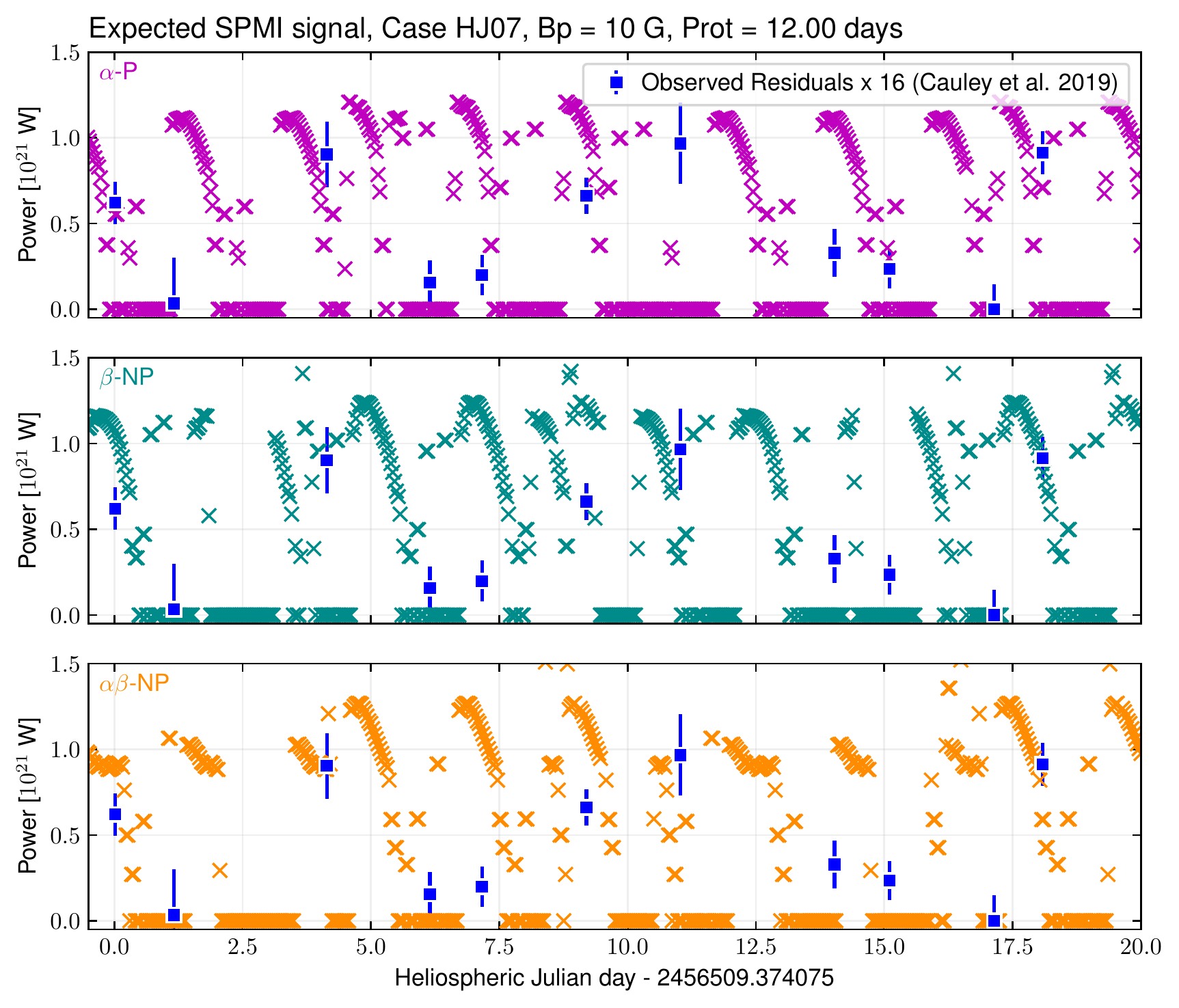}
  \caption{Predicted SPMI power as a function of time. Each panel corresponds to one magnetic extrapolation technique for the case HJ07, as indicated. The SPMI power is obtained assuming $B_P$=10 G. The SPMI power takes into account the orbital motion of the planet, the travel time between the planet and the star, the stellar rotation, and the inclination of the star with respect to the line of sight from Earth (\textit{i.e.} only the visible stellar disk is considered). Here, each cross represents the SPMI signal averaged over an exposure time of 1 hour to mimic the observed signal temporal resolution. More details are given in the text. The blue squares with error bars are the residuals of the power in the Ca II K band detected by \citet{Cauley2019}, multiplied by a factor 16 to ease the comparison with the predicted total SPMI power.}
  \label{fig:SPMIsignal}
\end{figure*}

We show the resulting theoretical SPMI power as a function of time in Fig. \ref{fig:SPMIsignal} for model HJ07. Each panel corresponds to a different extrapolation technique as indicated in the top left corners ($\alpha$-P, $\beta$-NP, and $\alpha\beta$-NP from top to bottom). The synthetic SPMI signal is shown with crosses and is averaged with a moving 1-hour window to mimic the observational exposure time \citep{Fares2017a}. We show here the total power involved in the SPMI estimated with the stretch-and-break mechanism ($\mathcal{P}_{\rm SB}$, see Eq. \ref{eq:PowerL13}), assuming a planetary magnetic field $B_P=10$ G. We note that the absolute value of the planetary magnetic field essentially changes the absolute power available (see Fig. \ref{fig:PowerAlongOrbit}) and affects only very mildly the shape of the synthetic SPMI signal. \moddAS{The inclination of the planetary field could nevertheless affect the strength of the SPMI, as was shown in \citet{Strugarek2015}. Here we consider the maximal interaction case, \textit{i.e.} the case where the polar planetary field is aligned with the ambient field.}

The three extrapolation models predict a significantly different SPMI signal. Models $\beta$-NP (middle panel, cyan) and $\alpha\beta$-NP (lower panel, orange) share some similarities but also present different structures, as visible for instance at 4 and 16 days. Model $\alpha$-P (top panel, purple) has features slightly shifted in time compared to the other two models, as for instance at 12 days. We therefore see here that the predicted SPMI signal strongly depends on the magnetic structure \moddAS{that sets the connectivity between the chromosphere and the planetary orbit}. 

We have added in each panel the residual power published by \citet{Cauley2019} shown as blue squares with error-bars. The residuals were obtained by subtracting a rotational modulation from the CaII emission (see \citet{Cauley2018,Cauley2019} for more details). \moddAS{Here, the predicted power is larger than the observed power in CaII by an order magnitude because we assumed a planetary field of 10 G (see Fig. \ref{fig:PowerAlongOrbit}). Indeed, only part of the total CaII power likely ends up in the CaII K band (see \S\ref{sec:conclusions} for discussions). Therefore, to ease the phase comparison with the synthetic SPMI signal, we have multiplied the residuals by 16.} \moddAS{The scarcity of the observational measures prevents us from performing a quantitative comparison of the observed and modelled signals. Qualitatively, we remark that model $\alpha$-P (top panel) predicts a lack of signal at epochs where the SPMI signal is strong, as for instance at about 11 days. Also, at about 14 days the model predicts a strong SPMI signal whereas the residual is very weak.} 
Models $\beta$-NP and $\alpha\beta$-NP reproduce well some observational epochs (\textit{e.g.} around 4 and 18 days), but fail at some other (\textit{e.g.} around 7 days). 

\refAS{To be more quantitative, we have performed a blind search of a periodic signal in the predicted SPMI signal following the same methodology of \citet{Cauley2018}. To do so, we sample the predicted signal with 11 points evenly distributed over the observational epoch of \citet{Cauley2018} and we randomly shift their location by at maximum 1 day. We also randomly generate a relative measurement error between 0 and 30\% for each point. We generate 1000 series of such synthetic observations. For each serie, we perfom a Lafler-Kinman statistic to search for a period, as in \citet{Cauley2018}. For that we use an open-source implementation of the Lafler-Kinman statistic\footnote{The \href{https://github.com/phuijse/P4J\#readme}{P4J} and the likelihood methodology described \href{https://github.com/phuijse/P4J/blob/master/examples/periodogram_demo.ipynb}{here}.}. We have first verified that this implementation applied on the original data of \citet{Cauley2019} (blue squares in Fig. \ref{fig:SPMIsignal}) does recover the orbital period of HD 189733b with a confidence level larger than 95\%, as in the original work of \citet{Cauley2018}. Applying the methodology on our sample of 1000 synthetic observations, we find that the orbital period of HD 189733b is recovered for 12\% to 23\% of the synthetic observations derived from the three HJ07 models, with a level of confidence larger than 95\%. This result is particularly interesting because \citet{Cauley2018} found a signal at the orbital period if HD 189733b for 1 out of 6 observational epochs, \textit{i.e.} a sucessful detection rate of about 17\%. It is therefore possible that the SPMI signal is actually present at all the epochs studied by \citet{Cauley2018}, but given the distribution of the observational data points it was most of the time unlikely to be detectable.}

Given the scarcity of the observational points, we nevertheless reiterate that it is not possible at this stage to confirm that the residuals can indeed be explained with a SPMI scenario. \refAS{The comparisons shown in the middle and bottom panels of Fig. \ref{fig:SPMIsignal} are yet encouraging, as well as the analysis of the detection likelihood we just detailed.} Much more intense observations, with multiple measurements within one night, could in principle capture the phase dynamics of the synthetic SPMI signal predicted here. We have no such observational coverage available so far in the literature. The results presented in this work therefore provide a pathway to firmly confirm SPMI interpretations of stellar activity residuals in compact exosystems. Such confirmation requires very regular and dedicated observations for a system such as HD 189733, \textit{e.g.} like the ambitious campaigns that are being carried with NenuFAR \citep{Zarka2020}.    



\section{Discussions and conclusions}
\label{sec:conclusions}

This paper is part of the MOVES collaboration, which is focused on characterizing the environment of HD 189733 and its hot Jupiter HD 189733b. It complements the initial MOVES I paper \citep{Fares2017a} that reconstructed the magnetic properties of HD 189733 over several epochs and the MOVES III paper \citep{Kavanagh2019} that modelled the wind of HD 189733 at three other epochs than the one studied here, with a focus on its impact of possible radio emissions.

In this paper (MOVES V), we have modelled the environment of HD 189733 for August 2013, based on its observed Zeeman-Doppler Imaging map \citep{Fares2017a}. At that epoch, \citet{Cauley2018} detected a signal in the CaII K band of HD 189733 that could originate from star-planet magnetic interaction. In this work, we assess the robustness of this interpretation based on a 3D modelling of the corona and wind of HD 189733.

We have used a 3D MHD approach with the framework WindPredict \citep{Reville2016a} based on the PLUTO code \citep{Mignone2007a}. This approach is similar to the approach that was followed in the MOVES III paper \citep{Kavanagh2019} using the BATS-R-US code. We studied twelve models covering four different thermodynamic coronal properties and three different magnetic extrapolation methods. The latter are an original development of this work, and their details are fully given in Appendix \ref{sec:Extrapol}. Three of the twelve models correspond to the same modelling choices as those made in \citet{Kavanagh2019}. The set of models presented in this work covers a large range of possible coronal states, and is available in the \href{http://www.galactica-simulations.eu/db/}{Open Database Galactica}. The existence of tracers of SPMI in the stellar signal requires that the planet HD 179733b orbits close enough to its host star, inside the Alf\'ven surface for at least part of its orbit. We find that this is possible only in half of our models using the relatively low coronal density and temperature. 

The set of models predicts the available Poynting flux at the planetary orbit. It is the primary energy source for SPMI regardless of the detailed model considered \citep{Zarka2001,Zarka2007,Saur2013,Lanza2013a,Strugarek2016c}. We find that the available Poynting flux varies from 10$^{-2}$ to about 4 W/m$^2$ in our models. The models predicting that HD 189773b is orbiting inside the Alfv\'en surface lead to a maximum Poynting flux of about 1 W/m$^2$ at the planetary orbit. We find that the available Poynting flux can vary by more than two orders of magnitude along the orbit of HD 189733b, and also by more than two orders of magnitude depending on the SPMI scenario. We find that the stretch-and-break mechanism of \citet{Lanza2013a} is the only mechanism providing enough power to tentatively explain the observations of \citet{Cauley2019}. 

The three magnetic field extrapolations considered in this work essentially affect the topology of the magnetized corona of HD 189733. We find that the amplitude of the SPMI is not changed significantly by the magnetic extrapolation.

We have assessed the temporal signature of the SPMI signal in the models where SPMI is able to affect the central star. \moddAS{We find that the magnetic extrapolation method completely determines the temporal phase of the predicted SPMI signal. At some specific epochs, some extrapolation methods predict a signal while others do not. \refAS{We have estimated that the detection of the orbital period in synthetic observations mimicking the sampling of \citet{Cauley2019} would be possible only from 12\% to 23\% of the time. This is compatible with the fact that \citet{Cauley2018} detected a signal on one observational epoch out of six.} 
We find that in the context of HD 189733, denser spectroscopic observations are needed to disambiguate the best extrapolation technique to reproduce the observational data, and are therefore needed to confirm the interpretation of excessive CaII K emissions as a result of SPMIs as was done in \citet{Cauley2019}. A denser dataset would also help identifying the stellar intrinsic variability associated to impulsive events such as flares as well as rotational variability due to the appearance and disappearance of plages and spots as the star rotates. \refAS{Indeed, flares and stellar activity can also be a source of noise hiding the SPMI signal.}} The predicted SPMI signal varies on a timescale of a few hours, therefore its firm detection requires several spectroscopic data points per night. In addition, the SPMI signal prediction requires constraints on the magnetic topology of HD 189733. The dense spectroscopic observations should therefore be complemented by concomitant and spectropolarimetric observations.

\moddAS{The estimated SPMI signal presented in this work takes into account the rotation of the star, the orbital phase of the planet, and the inclination of the exosytem. To obtain the ZDI map, \citet{Fares2017a} considered a differential rotation of ${\rm d}\Omega = 0.11 \pm 0.05$ rad.d$^{-1}$. We have not considered the effect of such a differential rotation on the SPMI connectivity over the timescale of a bit more than one stellar rotation. Coronal models taking into account the differential rotation \citep[\textit{e.g.}][]{Pinto2021} should be considered in future modelling efforts to characterize SPMIs in HD 189733.}

The strong variability of the SPMI signal in HD 189733 may explain why it was detected only during one epoch by \citet{Cauley2018}. We have shown that a SPMI signal is modulated by the orbital period of the planet, the rotation period of the star, and the magnetic topology linking the stellar surface to the planetary orbit. \refAS{This leads to a significant probability of non-detection of the SPMI signal when the observational data points are scarcely distributed.} Modelling in detail the connectivity and SPMI signal at other epochs of observations of HD 189733 could help to assess whether a SPMI signal could be buried in the existing observations, or if it is unlikely that SPMI was acting at that time. Again, a firm detection would still require much denser spectroscopic observations.  

We have shown that the interpretation of the observed stellar signal can lead to estimates of the planetary field up to values of about 30 G. This estimate includes the unknown conversion factor between the available power in the SPMI and the power in a given particular observational band. Future characterization of SPMI will require theoretical developments of the detailed physical mechanism behind energy deposition in the stellar atmosphere to put constraints on this conversion factor. In addition, the large planetary field leads to a large magnetosphere, as illustrated in Fig. \ref{fig:DummyPlanets}. If the planet indeed sustains such a large field, dedicated MHD models embedding a planet with is magnetosphere (\textit{e.g.} \citealt{Cohen2009b,Strugarek2015}) will be required to self-consistently estimate the power channeled by the SPMI and go beyond the scaling-law approach followed in this work. \moddAS{Such large magnetospheres could also present significant asymetries, which could affect their ionospheres and leave a detectable trace in phase curves of hot and ultra-hot Jupiters \citep[\textit{e.g.}][]{Helling2021}, and therefore offer a complementary view on these hypothetical planetary magnetospheres.}

\section*{Acknowledgements}

A. Strugarek thanks P.W. Cauley for sharing his data points shown in Fig. \ref{fig:SPMIsignal} \refAS{and for valuable suggestions that helped strengthen the conclusions of our work.} Computations were carried out using CEA TGCC and CNRS IDRIS facilities within the GENCI 60410133 and 80410133 allocations, and a local meso-computer founded by DIM ACAV+. A. Strugarek acknowledges funding from the European Union's Horizon-2020 research and innovation programme (Grant Agreement no. 776403 ExoplANETS-A), the PLATO/CNES grant at CEA/IRFU/DAp, and the Programme National de Plan\'etologie (PNP). A. Strugarek and A.S. Brun acknowledge funding from the ERC Synergy grant WholeSun 810218. R. Fares acknowledges funding from UAEU startup grant number G00003269. This work has been carried out in the frame of the National Centre for Competence in Research PlanetS supported by the Swiss National Science Foundation (SNSF). The authors acknowledge the financial support of the SNSF. This project has received funding from the European Research Council (ERC) under the European Union's Horizon 2020 research and innovation programme (project {\sc Spice Dune}, grant agreement No 947634). Ch.Helling acknowledges funding from the European Union H2020-MSCA-ITN-2019 under Grant Agreement no. 860470 (CHAMELEON). PJW acknowledges support from STFC through consolidated grants ST/L000733/1 and ST/P000495/1. AAV acknowledges funding from the European Research Council (ERC) under the European Union's Horizon 2020 research and innovation programme (grant agreement No 817540, ASTROFLOW). P. Zarka acknowledges funding from the European Research Council (ERC) under the European Union's Horizon 2020 research and innovation programme (grant agreement No 101020459 - Exoradio).

\section*{Data Availability}

The data presented in this work is made available to the community through the Open Database Galactica (\url{http://www.galactica-simulations.eu}). The 3D visualisation (Fig. \ref{fig:3Dexample}) and properties along the planet orbit (Fig. \ref{fig:PropAlongOrbit}) are downloadable for the 12 simulations on the \href{http://www.galactica-simulations.eu/db/STAR_PLANET_INT/HD189733_SPMI/}{Magnetic interactions in HD 189733} webpage. Request for accessing other data from the simulation can be addressed to \href{mailto:antoine.strugarek@cea.fr}{antoine.strugarek@cea.fr}.


\bibliographystyle{mnras}
\bibliography{library} 



\appendix

 



\section{Magnetic field Extrapolations}
\label{sec:Extrapol}

\subsection{Preamble}
\label{sec:preamble}

In this section, we give the analytical formulation for the three types of field extrapolation used in this study. They are easily derived using the vector spherical harmonics basis  
\begin{equation}
  \label{eq:RST}
  \left\{
  \begin{array}{lcl}
    \Rlm{} &=& \Ylm{} \er \\
    \Slm{} &=& \gradperp \Ylm{} = \dth\Ylm{}\ethe + \frac{1}{\sin{\theta}}\dphi\Ylm{}\ephi\\
    \Tlm{} &=& \gradperp\times\Rlm{} = \frac{1}{\sin{\theta}}\dphi\Ylm{}\ethe -\dth\Ylm{}\ephi
  \end{array}
  \right.
\end{equation}

Any field can be decomposed using this basis in 3D space, and we will use the following convention
\begin{equation}
  \mathbf{B} = \sum_{l=0}^{\infty} \sum_{m=-l}^{l} \alpha^{m}_{l} \mathbf{R}^{m}_{l} +
  \frac{\beta^{m}_{l}}{l+1}\mathbf{S}^{m}_{l} -
  \frac{\gamma^{m}_{l}}{l+1}\mathbf{T}^{m}_{l}\, ,
\end{equation}
where $\alpha^{m}_{l}$, $\beta^{m}_{l}$ and $\gamma^{m}_{l}$ are a function of $r$ only. \moddAS{In what follows we contract the summation symbol over $l$ and $m$ to $\sum_{l,m}$. The coefficients $\alpha^{m}_{l}$, $\beta^{m}_{l}$ and $\gamma^{m}_{l}$ correspond to the classical toroidal-poloidal decomposition of a divergence-free field, where $\alpha^{m}_{l}$ and $\beta^{m}_{l}$ carry the poloidal field and $\gamma^{m}_{l}$ carries the toroidal field.} Let us note upfront a few important relationships based on this decomposition:
  \begin{eqnarray}
    \label{eq:divergence}
    \Div\mathbf{B} &=&
    \sum_{l,m}\left[\frac{1}{r^2}\dr(r^2\alpha^{m}_{l})-l\frac{\beta^{m}_{l}}{r}\right]Y_l^m \\
    \label{eq:curl}
    \rot\mathbf{B} &=&
    \sum_{l,m}\left[-l\frac{\gamma^{m}_{l}}{r}\right]\Rlm{} +
    \left[-\frac{1}{r}\dr(r\gamma^{m}_{l})\right]\Slm{} \nonumber \\
        &+&
    \left[\frac{\alpha_l^m}{r}-\frac{1}{r(l+1)}\dr(r\beta^{m}_{l})\right]\Tlm{} \, , \\
\label{eq:integralB} 
    \iint {\bf B} \cdot {\bf B} {\rm d}\Omega &=& \sum_{l,m} \left|\alpha_l^m\right|^2 + \frac{l}{l+1}\left( \left|\beta_l^m\right|^2 + \left|\gamma_l^m\right|^2\right) \, ,
\end{eqnarray}
where ${\rm d}\Omega = \sin\theta{\rm d}\theta{\rm d}\varphi$. These formulae have some interesting implications. First, the $\gamma^m_l$ does not affect the divergence of $\mathbf{B}$, which means that we can choose any dependency we want for this coefficient without breaking $\Div{\bf B}=0$. Second, imposing $\Div{\bf B}=0$ gives a direct relationship between coefficients $\alpha^m_l$ and $\beta^m_l$. Finally, a current-free magnetic field requires that $\gamma^m_l(r)=0$, and sets another constraint on the relationship between $\alpha^m_l$ and $\beta^m_l$. We will now make use of these constraints to derive extrapolations methods. We will consider in all that follows that we have the knowledge of the decomposition coefficients on a spherical surface of radius $R_\star$, which is given by the ZDI technique. These coefficients are denoted $\alpha^\star_{lm}$, $\beta^\star_{lm}$ and $\gamma^\star_{lm}$.

\subsection{Potential extrapolation with a source surface ($\alpha$-P)}
\label{sec:potent-extr-1}

We first derive the potential extrapolation formulae based on a source
surface. The potential
assumption is $\bnab\times\mathbf{B}=0$, which directly translates to
$\mathbf{B} = -\bnab\phi$. Because $\bnab\cdot\mathbf{B}=0$,
$\phi$ is the solution of the Laplace equation which, using spherical
harmonics, can be written

\begin{equation*}
  \phi = \sum_{l,m} \left[\phi^{a}_{lm} r^{l}
    + \phi^{b}_{lm}r^{-(l+1)}\right] Y^{m}_{l} \, ,
\end{equation*}
hence
\begin{eqnarray}
  \mathbf{B}^{\alpha{\rm -P}} &=& \sum_{l,m} \left[-\phi^{a}_{lm} l r^{l-1}
    + \phi^{b}_{lm}(l+1)r^{-(l+2)}\right] \mathbf{R}^{m}_{l} \nonumber \\ &-& \left[
    \phi^{a}_{lm}r^{l-1} + \phi^{b}_{lm}r^{-(l+2)} \right]\mathbf{S}^{m}_{l} \, .
\label{eq:alphaP}
\end{eqnarray}

We suppose that $\mathbf{B}$ is purely radial outside a spherical
surface denoted $R_{ss}$, and known on the $R_\star$ spherical surface. As a consequence, we can write 
\begin{equation*}
  \left\{
    \begin{array}{ccc}
      -\phi^{a}_{lm} l R_\star^{l-1}
    + \phi^{b}_{lm}(l+1)R_\star^{-(l+2)} & = & \alpha^{\star}_{lm} \\
    \phi^{a}_{lm}R_{ss}^{l-1} + \phi^{b}_{lm}R_{ss}^{-(l+2)} &=& 0
    \end{array}
  \right.
\end{equation*}
which leads to
\begin{equation*}
  \left\{
    \begin{array}{ccc}
      \phi^{b}_{lm} &=& \frac{\alpha^{\star}_{lm}}{lR_\star^{l-1}R_{ss}^{-(2l+1)} +
        (l+1)R_\star^{-(l+2)}} \\
      \phi^{a}_{lm} &=& -\phi^{b}_{lm}R_{ss}^{-(2l+1)}
    \end{array}
  \right. \, .
\end{equation*}

Hence, using the decomposition given in Section
\ref{sec:preamble}, we have for the potential field:
\begin{eqnarray*}
  \alpha_{lm}^{\alpha{\rm -P}} &=& \alpha^{\star}_{lm} \frac{ l
    (R_\star/R_{ss})^{2l+1} (r/R_\star)^{l-1} +
    (l+1)(r/R_\star)^{-(l+2)}}{l(R_\star/R_{ss})^{2l+1} + (l+1)} \, ,\\
  \beta_{lm}^{\alpha{\rm -P}} &=& \left(l+1\right)\alpha^{\star}_{lm} \frac{ 
    (R_\star/R_{ss})^{2l+1} (r/R_\star)^{l-1} -
    (r/R_\star)^{-(l+2)}}{l(R_\star/R_{ss})^{2l+1} + (l+1)} \, ,\\
  \gamma_{lm}^{\alpha{\rm -P}} &=& 0 \, .\\
\end{eqnarray*}

This extrapolation leads to a field that matches only the radial component of ${\bf B}$ on $R_\star$, and is denoted $\alpha$-P in this work. It is illustrated in the second column of Fig. \ref{fig:MagMap}. 

\subsection{Non-potential extrapolation matching the horizontal field ($\beta$-NP)}
\label{sec:non-potent-extr}

We now present an extrapolation method that allows to match the horizontal field $B_\theta {\bf e}_\theta + B_\varphi {\bf e}_\varphi$ on surface $R_\star$. This extrapolation is necessarily partly non-potential, as any potential field requires $\gamma^m_l(r)=0$ as we have seen. 

We first derive a potential part in the same spirit as \S\ref{sec:potent-extr-1}, but based on matching $\beta^m_l$ at $R_\star$ this time. Starting from Eq. \ref{eq:alphaP} we now obtain that
\begin{equation*}
  \left\{
    \begin{array}{ccc}
      \phi^{a}_{lm} R_\star^{l-1}
    + \phi^{b}_{lm}R_\star^{-(l+2)} & = & -\frac{\beta^\star_{lm}}{l+1} \\
    \phi^{a}_{lm}R_{ss}^{l-1} + \phi^{b}_{lm}R_{ss}^{-(l+2)} &=& 0
    \end{array}
  \right.
\end{equation*}
which leads to
\begin{equation*}
  \left\{
    \begin{array}{ccc}
      \phi^{b}_{lm} &=& -\frac{\beta^\star_{lm} R_\star^{l+2}}{l+1} \left[1-\left(\frac{R_\star}{R_{ss}}\right)^{2l+1}\right]^{-1} \\
      \phi^{a}_{lm} &=& -\phi^{b}_{lm}R_{ss}^{-(2l+1)}
    \end{array}
  \right.\, , 
\end{equation*}
and finally gives
\begin{eqnarray*}
  \alpha_{lm}^{\beta{\rm -NP}} &=& \beta^\star_{lm} \left(\frac{R_\star}{R_{ss}}\right)^{l+2} \frac{ l/(l+1) \left(r/R_{ss}\right)^{2l+1} + 1  }{\left(R_\star/R_{ss}\right)^{2l+1}-1}  \, ,\\
  \beta_{lm}^{\beta{\rm -NP}} &=& \beta^\star_{lm} \left(\frac{R_\star}{R_{ss}}\right)^{l+2} \frac{ \left(r/R_{ss}\right)^{2l+1} - 1  }{\left(R_\star/R_{ss}\right)^{2l+1}-1} \, .
\end{eqnarray*}
    
To complete this extrapolation we must now specify $\gamma^{\beta{\rm -NP}}_{lm}(r)$. The only strong constraint is that it must match $\gamma^\star_{lm}$ on $R_\star$, since it does not affect the divergence of ${\bf B}$. Here we have chosen a simple power law defined as 
\begin{equation*}
    \gamma^{\beta{\rm -NP}}_{lm} = \gamma^\star_{lm}\frac{r^{-n_\gamma}-R_{ss}^{-n_\gamma}}{R_\star^{-n_\gamma}-R_{ss}^{-n_\gamma}}\, .
\end{equation*}
In this work we chose $n_\gamma=5$ to ensure that $\gamma_{lm}(r)$ quickly decreases when compared to $\alpha_{lm}$ and $\beta_{lm}$. \moddAS{Note that $\gamma_{lm}^{\beta{\rm -NP}}$ carries the free magnetic energy of the extrapolated field. A smaller value of $n_\gamma$ would lead to a higher free energy in the initial volume.}

This extrapolation matched exactly the horizontal field on $R_\star$, and is illustrated on the third column of Figure \ref{fig:MagMap}.

\subsection{Mixed non-potential extrapolation ($\alpha\beta$-NP)}
\label{sec:non-potent-extr-mixt}

In this work we make use of a third extrapolation method dubbed $\alpha\beta$-NP. The complete extrapolation leveraging a 3D surface field is complex \citep[\textit{e.g.}][]{Amari2013} and requires the development of dedicated codes (for a review in the context of the solar corona and wind, see \citealt{Yeates2018}). Such techniques are out of the scope of the present work, and we opted for a simple linear combination between $\alpha$-P and $\beta$-NP extrapolations to approximate an extrapolation matching the three components of ${\bf B}$ on $R_\star$. This simply writes 
\begin{align*}
    \alpha_{lm}^{\alpha\beta{\rm -NP}} =& g_{lm}\alpha_{lm}^{\alpha{\rm -P}}+h_{lm}\alpha_{lm}^{\beta{\rm -NP}} \, ,\\
    \beta_{lm}^{\alpha\beta{\rm -NP}} =& g_{lm}\beta_{lm}^{\alpha{\rm -P}}+h_{lm}\beta_{lm}^{\beta{\rm -NP}} \, ,\\
    \gamma_{lm}^{\alpha\beta{\rm -NP}} =& g_{lm}\gamma_{lm}^{\alpha{\rm -P}}+h_{lm}\gamma_{lm}^{\beta{\rm -NP}}\, .\\
\end{align*}

This extrapolation approximately matches the three components of ${\bf B}$ on $R_\star$ and is illustrated in the last column of Fig. \ref{fig:MagMap}. It is a harmonic-by-harmonic linear combination of the $\alpha$-P and $\alpha\beta$-NP extrapolation and therefore always satisfies $\boldsymbol{\nabla}\cdot{\bf B}=0$. The coefficients $g_{lm}$ and $h_{lm}$ can be chosen to match better one of the components, and can be tailored to maximize the closeness of the reconstructed surface field with the ZDI map. In this work, we have chosen the simplest linear combination $g_{lm}=0.5$ and $h_{lm}=0.5$ which achieves a satisfying reconstruction since the toroidal energy of the ZDI magnetic field is roughly 50\% of the total surface field energy (see first line, second column in Table \ref{ta:Emodels}). 

\section{Star-planet magnetic interaction and stellar rotation period}
\label{sec:Prot_sensitivity}

\modAS{
The SPMI signal shown in Fig. \ref{fig:SPMIsignal} depends on stellar and planetary parameters. The amplitude of the signal depends on $B_P$ (as shown in Fig. \ref{fig:PowerAlongOrbit}). The shape of the signal depends on the orbital parameters (semi-major axis, orbital period) which are well constrained by observations (see Table \ref{ta:HD189733prop}). The uncertainties on the orbital parameters are too small to influence our results. Likewise, the inclination of the system can in principle change significantly the SPMI signal, as was shown in Appendix A of \citet{Strugarek2019}. In the case of HD 189733 \citet{Fares2017a} used an inclination of $5^\circ$ with respect to the axis perpendicular to the line of sight, and \citet{Cegla2016} obtained an inclination of $10^\circ$. We have tested both inclinations and found no significant difference in our results. }

\begin{figure*}
  \centering
    \includegraphics[width=\linewidth]{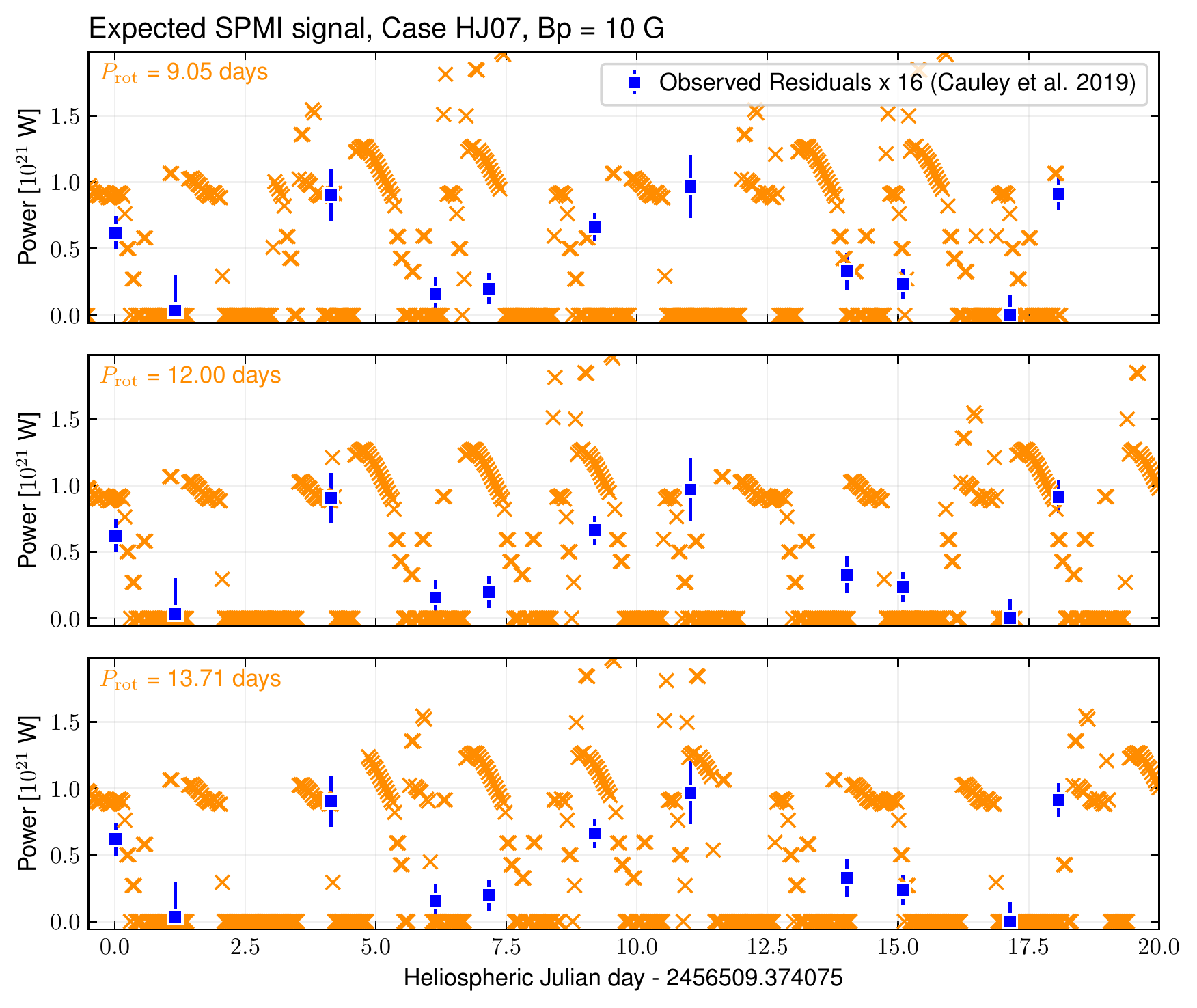}
  \caption{Predicted SPMI power as a function of time, for model HJ07 $\alpha\beta-NP$. The layout is the same as Fig. \ref{fig:SPMIsignal}, except that here each panel correspond to a different stellar rotation period, taken from \citet{Cegla2016}.}
  \label{fig:Prot_sensitivity}
\end{figure*}

\modAS{The rotation period of HD 189733 is nevertheless directly determining the phase of the SPMI signal we predict. HD 189733 possesses a surface differential rotation \citep[\textit{e.g.}][]{Fares2010a}. Its equatorial rotation rate has been characterized by multiplied studies. \citet{Cegla2016} compiled the different values published at that time, and derived yet another rotation period based on the Rossiter-McLaughlin effect. The published equatorial rotation rates therefore vary from 9.05 days \citep{Cegla2016} to 13.71 days \citep{Winn2006}. We have covered this range here to assess the influence of the uncertainty on the equatorial rotation period of HD 189733 on the predicted SPMI signal. The results are shown in Fig. \ref{fig:Prot_sensitivity} with the same layout as Fig. \ref{fig:SPMIsignal} and only for model HJ07 ($\alpha\beta$-NP). The temporal signature of the SPMI is strongly influenced by $P_{\rm rot}$, as expected. For instance, in the fast-rotating case (top panel), a strong long-lasting signal is expected near 10 days whereas it is absent in the two other cases rotating slower. The rotation period of the star varies by 14\% between the middle panel ($P_{\rm rot}$=12 days) and lower panel ($P_{\rm rot}$=13.71 days). In this case, the SPMI signal is comparable but can still change significantly at some specific epochs like around 18 days, where the middle panel predicts a strong SPMI signal and the lower panel does not. We conclude here that the prediction of SPMI requires a careful characterization of the stellar rotation rate.
}

\section{Star-planet magnetic interaction and propagation time}
\label{sec:Propagation sensivity}

\begin{figure*}
  \centering
    \includegraphics[width=\linewidth]{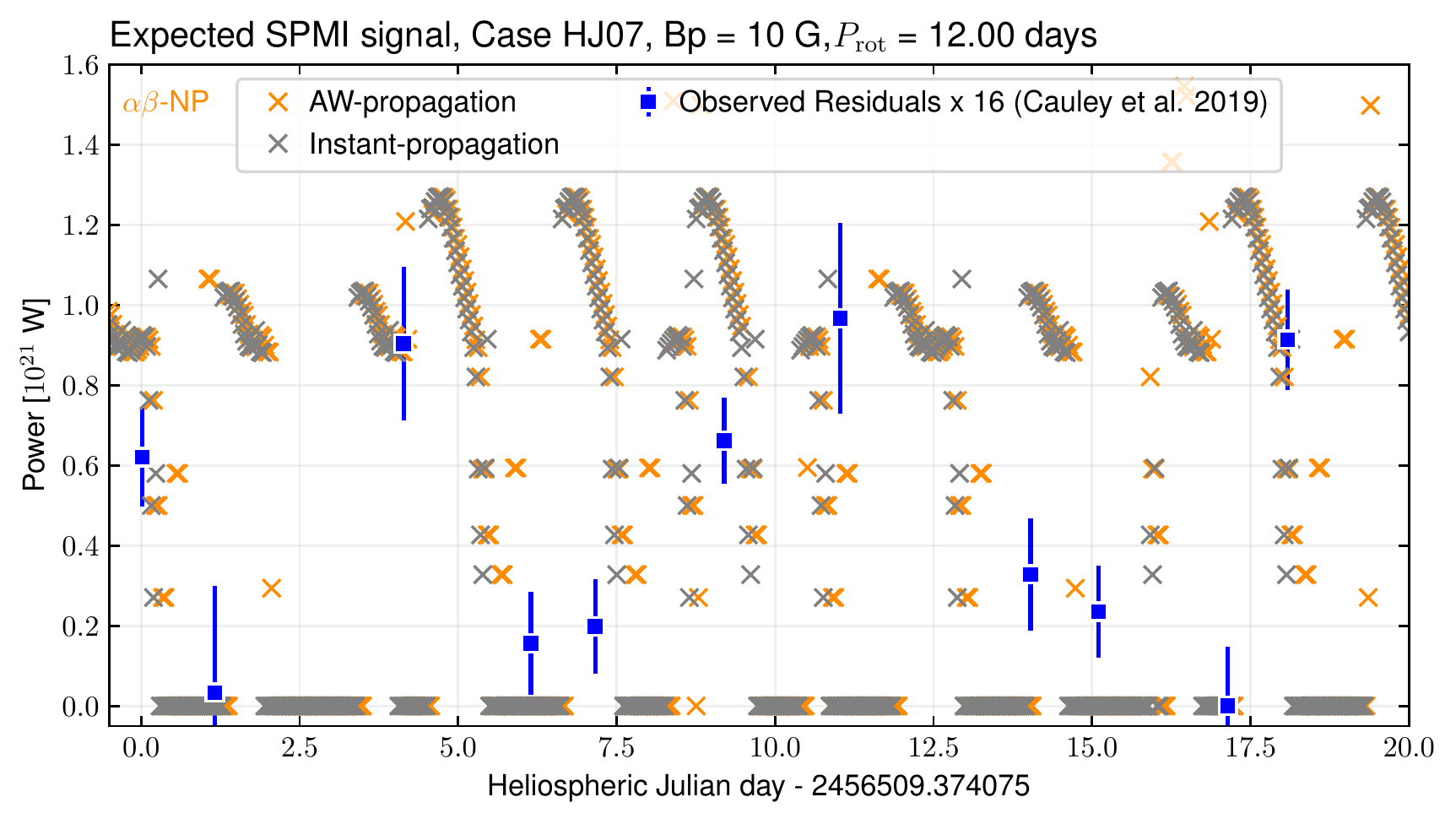}
  \caption{Predicted SPMI power as a function of time. The layout is the same as Fig. \ref{fig:SPMIsignal}. The orange crosses correspond to the case where the SPMI signal is carried by Alfv\'en waves from the vicinity of the planet down to the chromosphere, and the gray crosses to the case where it is carried by relativistic electrons accelerated in the vicinity of the planet.}
  \label{fig:Prop_sensitivity}
\end{figure*}

\moddAS{
The SPMI signal can be transported from the planet to the star by several means. The canonical Alfv\'en-wing scenario \citep{Saur2013} considers Alfv\'en waves that propagate at the local Alfv\'en speed from the vicinity of the planet down to the low corona, where they may deposit energy and accelerate electrons leading to an observable signal. Nevertheless, relativistic electrons can also be accelerated at magnetic reconnection sites close to the planet. In the latter case, the SPMI signal is transported very fast from the planet vicinity to the chromosphere. To assess the effect of the two scenarios on the predicted SPMI signal, we show them in Fig. \ref{fig:Prop_sensitivity}. The orange crosses are the canonical case of a transport by Alfv\'en waves that we considered in this work. The gray crosses correspond to an instantaneous travel time between the planet and the chromosphere, mimicking the transport by relativistic electrons. We see that the bulk part of the SPMI signal is only slightly shifted from one scenario to the other, because the local Alfv\'en crossing time is fast for most of the orbit (typically of the order of about an hour). At some specific epochs, for instance near 6 days, the Alfv\'en crossing time can reach up to 35 hours and the two signals start to differ significantly. In the context of HD 189733 in August 2013, the orbital phased with a long Alfv\'en crossing times are not abundant and therefore the two scenarios are indistinguishable with the available observational data.  
}

\bsp	
\label{lastpage}
\end{document}